\documentclass[a4paper,11pt]{article}
\pdfoutput=1
\usepackage{jheppub}
\usepackage{graphicx}
\usepackage{subcaption}
\usepackage{dsfont}
\usepackage{etoolbox}
\makeatletter
\patchcmd{\maketitle}{\@fpheader}{\\}{}{}
\makeatother

\title{A perturbative perspective on self-supporting wormholes}

\author[a]{Zicao Fu,}
\author[a]{Brianna Grado-White,}
\author[a]{and Donald Marolf}

\affiliation[a]{Department of Physics, University of California, Santa Barbara, CA 93106, USA}

\emailAdd{zicaofu@physics.ucsb.edu}
\emailAdd{brianna@physics.ucsb.edu}
\emailAdd{marolf@physics.ucsb.edu}

\abstract{We describe a class of wormholes that generically become traversable after incorporating gravitational back-reaction from linear quantum fields satisfying appropriate (periodic or anti-periodic) boundary conditions around a non-contractible cycle, but with natural boundary conditions at infinity (i.e., without additional boundary interactions).  The class includes both asymptotically flat and asymptotically AdS examples.   Related constructions can also be performed in asymptotically de Sitter space or in other closed cosmologies. Simple asymptotically AdS$_3$ or asymptotically AdS$_3 \times S^1$ examples with a single periodic scalar field are then studied in detail.  When the examples admit a smooth extremal limit, our perturbative analysis indicates the back-reacted wormhole remains traversable at later and later times as this limit is approached.  This suggests that a fully non-perturbative treatment would find a self-supporting eternal traversable wormhole. While the general case remains to be analyzed in detail, the likely relation of the above effect to other known instabilities of extreme black holes may make the construction of eternal traversable wormholes more straightforward than previously expected.}

\begin{document}
\maketitle

\section{Introduction}

Wormholes have long been of interest to both scientists (see e.g. \cite{Einstein:1935tc,Graves:1960zz,Morris:1988tu}) and the general public, especially in the context of their possible use for rapid transit or communication over long distances.  While the topological censorship theorems \cite{Friedman:1993ty,Galloway:1999bp} forbid traversable wormholes in Einstein-Hilbert gravity coupled to matter satisfying the null energy condition (NEC) $T_{ab} k^a k^b \ge 0$, the fact that quantum fields can violate the NEC (and that higher-derivative corrections can alter the dynamics away from Einstein-Hilbert) has led to speculation (e.g. \cite{Morris:1988tu}) that traversable wormholes might nevertheless be constructed by sufficiently advanced civilizations.

Indeed, an Einstein-Hilbert traversable wormhole supported by quantum fields was recently constructed in \cite{Gao:2016bin}. Their wormhole connects two asymptotically 2+1-dimensional anti-de Sitter (AdS) regions that are otherwise disconnected in the bulk spacetime.  However, the model contains an explicit non-geometric time-dependent coupling of quantum field degrees of freedom near one AdS boundary to similar degrees of freedom near the other. Turning on this coupling briefly near $t=0$ allows causal curves that begin at one AdS boundary in the far past to traverse the wormhole and reach the other boundary in some finite time.  Though the wormhole collapses and becomes non-traversable at later times, the negative energy induced by the boundary coupling supports a transient traversable wormhole. The extension to the rotating case was performed in \cite{Caceres:2018ehr}.

Here and below, we define the term ``traversable wormhole'' to mean a violation of the topological censorship results of \cite{Friedman:1993ty,Galloway:1999bp}; i.e., it  represents causal curves that cannot be deformed (while remaining causal) to lie entirely in the boundary of the given spacetime. Note that there exist interesting solutions of Einstein-Hilbert gravity involving thin necks connecting large regions (e.g. \cite{Bachas:2017rch}) which are not wormholes in this sense. In addition, an analogue of the effect in \cite{Gao:2016bin} without wormholes was recently discussed in \cite{Almheiri:2018ijj}.

\begin{figure}[t]
        \begin{center}
                \includegraphics[width=0.32\textwidth]{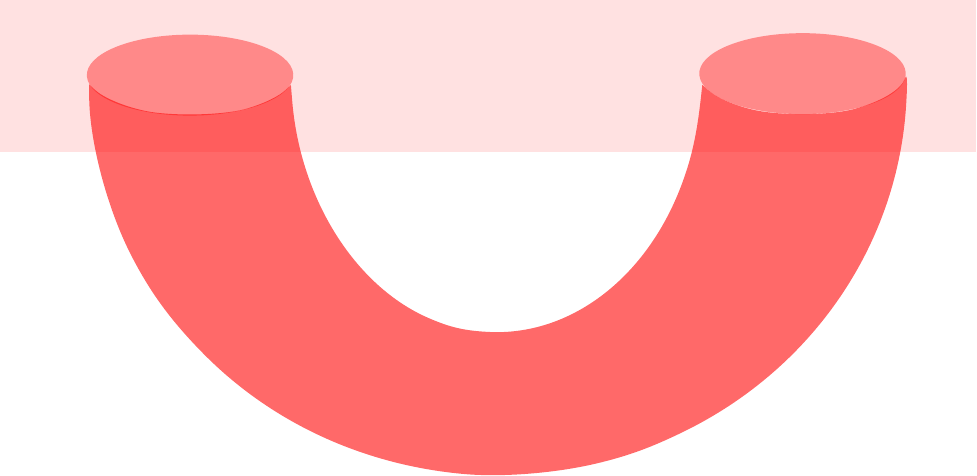}
        \end{center}
        \caption{A moment of time in a spacetime with a wormhole (shaded region) formed by adding a handle to a space with a single asymptotic region.}
        \label{fig:wormhole}
\end{figure}

From the perspective of the bulk spacetime, boundary interactions like those used in \cite{Gao:2016bin} are both non-local and acausal. However, it is expected that similar boundary couplings can be induced by starting with local causal dynamics on a spacetime of the form described by figure \ref{fig:wormhole}, in which the ends of the wormhole interact causally through the ambient spacetime.  Integrating out the unshaded region in figure \ref{fig:wormhole} clearly leads to an interaction between opposite ends of the wormhole (shaded region).  Though not precisely of the form studied in \cite{Gao:2016bin}, the details of the boundary coupling do not appear to be critical to the construction.

Indeed, during the final preparation of this manuscript, a traversable wormhole was constructed \cite{Maldacena:2018gjk} using only local and causal bulk dynamics.  In addition, the wormhole of \cite{Maldacena:2018lmt} is stable and remains open forever.  We refer to such wormholes as self-supporting.  This construction was inspired by \cite{Maldacena:2018lmt}, which showed that adding time-independent boundary interactions to AdS$_2$ in some cases leads to static (eternal) traversable wormholes that in particular are traversable at any time.  In \cite{Maldacena:2018lmt}, the eternal wormholes arise as ground states and, as the authors of \cite{Maldacena:2018lmt} point out, more generally the time-translation invariance of a ground state leads one to expect that a geometric description must have either a static traversable wormhole or no wormhole at all\footnote{Recall that familiar non-traversable wormholes like Reissner-Nordstr\"om, Kerr, or BTZ degenerate and disconnect in the limit of zero temperature.  The full argument is best given in Euclidean signature so as to exclude non-traversable static wormholes of the form discussed in \cite{Fu:2016xaa}.  This is appropriate for a ground state defined by a Euclidean path integral. \label{foot:static}}.

The wormholes constructed in \cite{Maldacena:2018lmt} are extremely fragile, yet in some sense their construction was easier than had long been assumed.  It is therefore useful to find a clean and simple perspective explaining why self-supporting wormholes should exist.  We provide such an explanation below using first-order perturbation theory about classical solutions.  Indeed, we will find perturbative indications that self-supporting wormholes can indeed exist even when the number of propagating quantum fields is small.  Our examples resemble the $\Delta < 1/2$ case studied in \cite{Maldacena:2018lmt} in that the back-reaction grows in the IR limit.   While by definition there can be no traverseable wormholes of our sort in closed cosmologies, there can nevertheless be related effects.  For example, one can use these techniques to build a Schwarzschild-de Sitter-like solution in which causal curves from $I^-$ to $I^+$ can pass through the associated Einstein-Rosen-like bridge.

\begin{figure}[t]
\begin{subfigure}{.49\linewidth}
\centering
\includegraphics[width=1\linewidth]{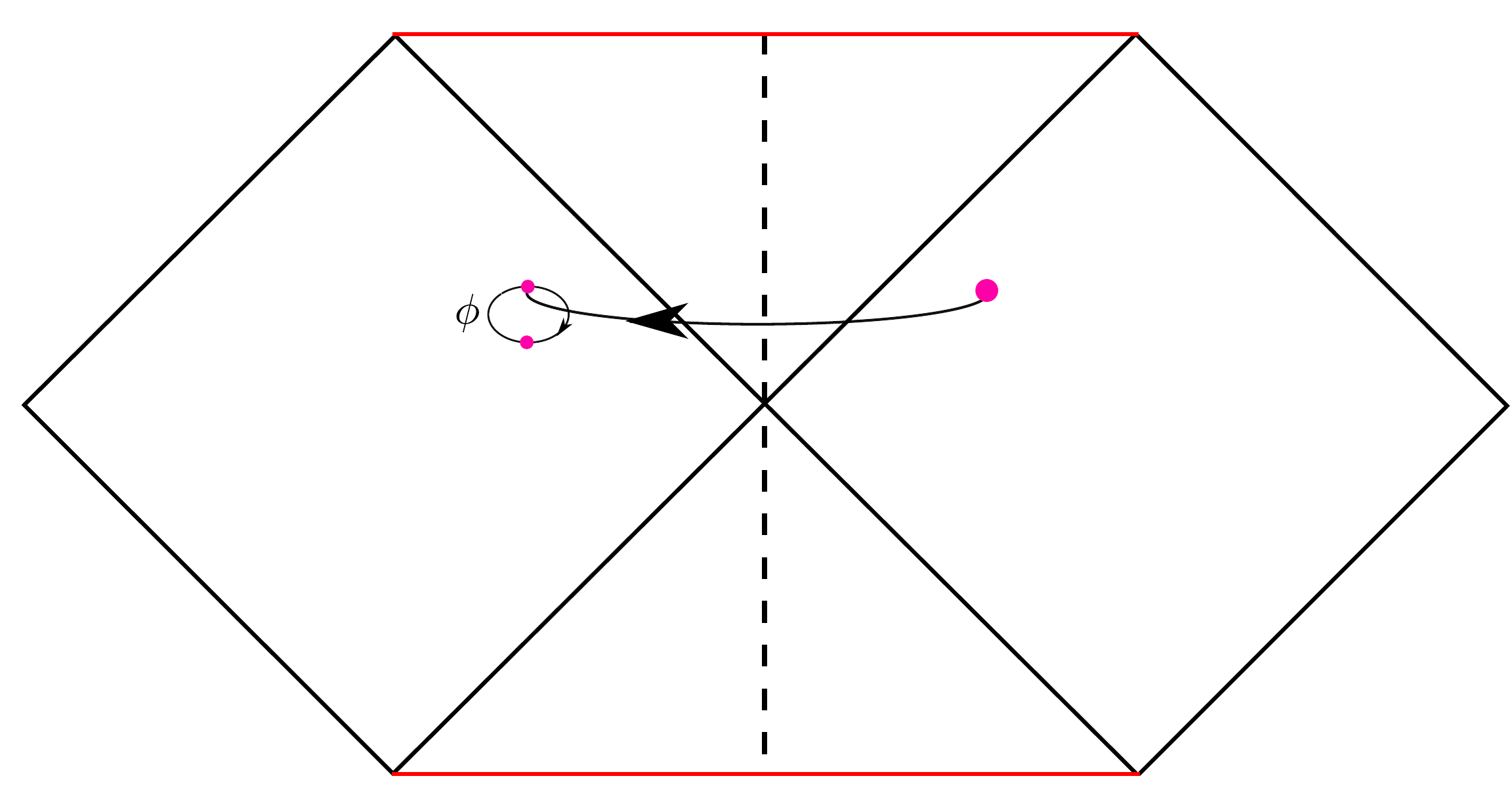}\quad%
\end{subfigure}
\begin{subfigure}{.49\linewidth}
\centering
\includegraphics[width=0.49\linewidth]{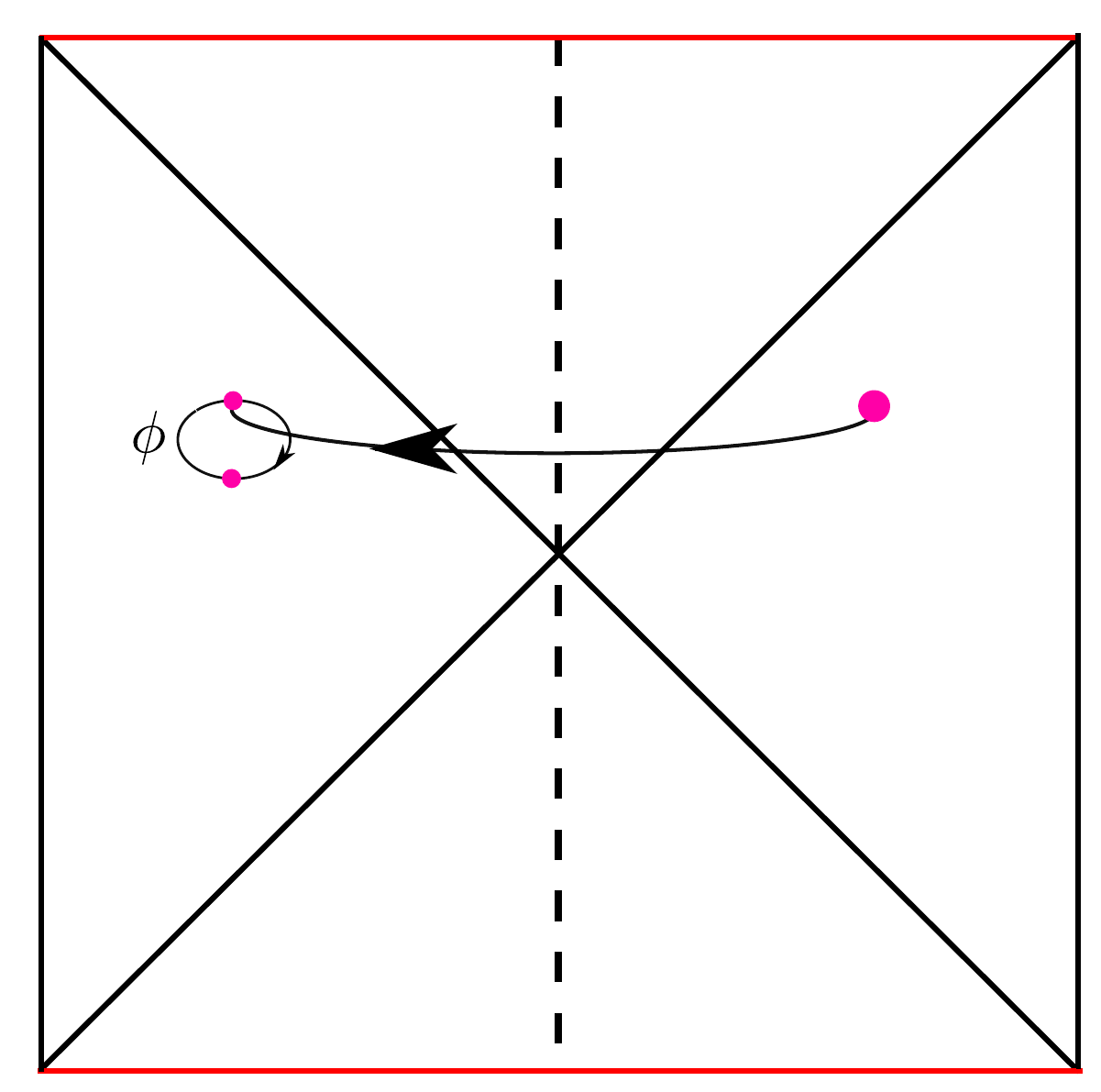}\quad%
\end{subfigure}
\caption{The asymptotically flat $\mathbb{RP}^3$ geon and the AdS $\mathbb{RP}^2$ geon are respectively ${\mathbb Z}_2$ quotients of Kruskal's extension (left) of the Schwarzschild solution and BTZ (right).  The quotients act on the above conformal diagrams by reflection across the dashed lines, and simultaneously act as the antipodal map (see insets) on the suppressed $S^2$ or $S^1$. Due to this combined action, the resulting spacetimes are smooth.  However, since this action maps the Killing field $\xi^a$ to $-\xi^a$, the geon quotients lack globally-defined time-translation Killing fields. In particular, the dashed lines are orthogonal to preferred spacelike surfaces of vanishing extrinsic curvature that one may call $t=0$. Our Kaluza-Klein end-of-the-world brane is a quotient of BTZ $\times S^1$ by a related isometry that acts trivially on the BTZ $\phi$-circle but acts on the internal $S^1$ via the antipodal map.}
        \label{fig:geons}
\end{figure}

At least for the purpose of establishing transient traversability for some choice of boundary conditions, the important properties of our backgrounds are that they are smooth, globally hyperbolic ${\mathbb Z}_2$ quotients of spacetimes with bifurcate Killing horizons and well-defined Hartle-Hawking states under an isometry that exchanges the left- and right-moving horizons.  Such spacetimes may be said to generalize the $\mathbb{RP}^3$ geon described in \cite{GiuliniPhD,Friedman:1993ty} (and in \cite{Misner:1957mt} at the level of time-symmetric initial data); see figure \ref{fig:geons} (left).    However,  as discussed in section \ref{sec:disc}, they may also take the more familiar form shown in figure \ref{fig:wormhole}.

Much like the Ba\~nados-Teitelboim-Zanelli (BTZ) case studied in \cite{Gao:2016bin}, the bifurcate horizon in the covering space makes the wormholes nearly traversable, so that they might be rendered traversable by the perturbatively small backreaction sources of a quantum field.  On the other hand, linear quantum fields in backgrounds with global Killing symmetries satisfy the averaged null energy condition (ANEC), meaning that the integral of $T_{ab}k^ak^b$ over complete null generators is non-negative\footnote{This follows for both free and super-renormalizeable field theories from e.g. combining the results of \cite{Wall:2009wi} with those of \cite{Wall:2011hj}, or from the free-field quantum null energy condition (QNEC) derived in \cite{Bousso:2015wca}.   This result should also hold for quantum field theories that approach a non-trivial UV conformal fixed point as one expects that the arguments of \cite{Faulkner:2016mzt,Hartman:2016lgu} generalize (at least in the static case where analytic continuation is straightforward) directly to Killing horizons in curved spacetimes.  For such more general theories, one could alternately use the QNEC connection of \cite{Bousso:2015wca} and generalize the results of \cite{Balakrishnan:2017bjg} to appropriate Killing horizons.}.  Thus, a bifurcate Killing horizon will not become traversable under first-order back-reaction from quantum fields in any quantum state and the above-mentioned ${\mathbb Z}_2$-quotient operation plays a key role in our analysis below.

After describing the general framework for such constructions and the relation to \cite{Gao:2016bin} and \cite{Maldacena:2018lmt} in section \ref{sec:general},
we study simple examples of transient such traversable wormholes
in section \ref{sec:static}, and a more complicated example in \ref{sec:KKZBO} that admits an extremal limit in which the wormhole appears to remain open forever.   We consider only scalar quantum fields in the work below, though similar effects should be expected from higher spin fields.  It would be particularly interesting to study effects from linearized gravitons.

For simplicity, section \ref{sec:static} considers backgrounds defined by the AdS $\mathbb{RP}^2$ geon \cite{Louko:1998hc} and simple Kaluza-Klein end-of-the-world branes\footnote{Such spacetimes (for AdS$_d$ with $d\ge 4$) were studied in \cite{Louko:2004ej} where they were called higher-dimensional geons.  We use the term KKEOW brane here as we emphasize the AdS$_3$ perspective, and in particular because it provides a smooth top-down model of the end-of-the-world brane spacetime of \cite{Hartman:2013qma,Almheiri:2018ijj}.}  (KKEOW branes) that are respectively quotients of AdS$_3$ and AdS$_3 \times S^1$.  In particular, the former are ${\mathbb Z}_2$ quotients of BTZ spacetimes, and the latter are quotients of BTZ $\times S^1$; see figure \ref{fig:geons}.
In each case, as explained in section \ref{sec:general}, we take the bulk quantum fields to be in the associated Hartle-Hawking state defined by the method of images using the above ${\mathbb Z}_2$ quotient, or equivalently defined via a path integral over the Euclidean section of the background geometry. Both backgrounds define wormholes with ${\mathbb Z}_2$ homotopy\footnote{\label{foot:wpi} It is useful to define the wormhole homotopy group to be the quotient $\pi_1^w{M} := \pi_1(M)/\pi_1(\partial M)$ of the bulk homotopy group $\pi_1(M)$ by the boundary homotopy group $\pi_1(\partial M)$.  If there is no boundary $(\partial M = \emptyset)$, we define $\pi_1(\emptyset)$ to be trivial. The examples described here have $\pi_1^w(M) = {\mathbb Z}_2$. } fully hidden by a single black hole horizon.  They are also non-orientable, though with additional Kaluza-Klein dimensions they admit orientable cousins as in \cite{Louko:1998hc}.  Outside the horizons, the spacetimes are precisely BTZ or BTZ $\times S^1$, and even inside the horizon these quotients preserve exact rotational symmetry.

Unfortunately, the examples of section \ref{sec:static} do not admit smooth zero-temperature limits.  We thus turn in section \ref{sec:KKZBO} to a slightly more complicated ${\mathbb Z}_2$ quotient of BTZ $\times S^1$ that breaks rotational symmetry but nevertheless supports the addition of angular momentum.  The four-dimensional spacetime is smooth, though after Kaluza-Klein reduction on the $S^1$, the resulting three-dimensional spacetime has two conical singularities with $\pi$ deficit angles.  We therefore refer to this example as describing Kaluza-Klein zero-brane orbifolds (KKZBOs).  This construction admits a smooth extremal limit and (as it turns out) yields an orientable spacetime.
In our first-order perturbative analysis, back-reaction renders the KKZBO wormhole traversable until a time $t_f$ that becomes later and later as extremality is approached.  This suggests that a complete non-perturbative analysis would find a self-supporting eternal traversable wormhole.  The large effect near extremality is associated with a divergence of the relevant Green's function in the extremal limit.  It would be interesting to better understand the relationship of this divergence to other known instabilities of extreme black holes.

We end with some discussion in section \ref{sec:disc}, focusing on back-reaction in the extremal limit, showing that the general class of wormholes described in section \ref{sec:general} includes wormholes of the familiar form depicted in figure \ref{fig:wormhole}.  In particular, assuming that perturbations of Reissner-Nordstr\"om black holes display an instability similar to the one noted above for extreme BTZ, our mechanism also appears to explain the existence of the self-supporting wormholes constructed in \cite{Maldacena:2018gjk}.  An appendix also describes a slight generalization of the framework from section \ref{sec:general}.

\section{${\mathbb Z}_2$-quotient wormholes and their Hartle-Hawking states}
\label{sec:general}

As stated above, at least for the purpose of establishing transient traversability, the important properties of our backgrounds $M$ are that they are smooth globally hyperbolic ${\mathbb Z}_2$ quotients of spacetimes $\tilde M$ with bifurcate Killing horizons and well-defined Hartle-Hawking states $|0_{HH, \tilde M} \rangle$ under a discrete ${\mathbb Z}_2$ isometry $J$ (i.e., with $J^2 = \mathds{1}$) that exchanges the left- and right-moving horizons.  Here, by a Hartle-Hawking state we mean a state of the quantum fields that is smooth on the full bifurcate horizon and invariant under the Killing symmetry. Such spacetimes $M$ are then generalizations of the (Schwarzschild) $\mathbb{RP}^3$ geon described in \cite{GiuliniPhD,Friedman:1993ty} and the $\mathbb{RP}^2$ AdS geon of \cite{Louko:1998hc}; see figure \ref{fig:geons}. For later purposes, note that the homotopy group $\pi_1(\tilde M)$ is a normal subgroup of $\pi_1(M)$ with $\pi_1(M)/\pi_1(\tilde M) = \mathbb{Z}_2$.  In order to describe the additional topology introduced by the $\mathbb{Z}_2$ quotient, it will be useful to choose some associated $\gamma \in \pi(M)$ which projects to the non-trivial element of $\pi_1(M)/\pi_1(\tilde M) = \mathbb{Z}_2$ and for which $\gamma^2 = \mathds{1}$.

To set the stage for detailed calculations in section \ref{sec:static}, we give a simple argument in section \ref{sec:free} below that -- for either periodic or anti-periodic boundary conditions around $\gamma$ -- this setting generically leads to traversability in the presence of free quantum fields.  In order to provide a useful perspective and explore connections with both recent work \cite{Maldacena:2018lmt} by Maldacena and Qi and the original traversable wormhole \cite{Gao:2016bin} of Gao, Jafferis, and Wall, section \ref{sec:PI} then describes an alternate construction via path integrals that generalizes this construction to interacting fields.

\subsection{The free field case}
\label{sec:free}

Our explicit work in sections \ref{sec:static} and \ref{sec:KKZBO} below involves free quantum fields.  We may therefore follow  \cite{Louko:1998dj,Louko:1998hc} and define a state on the quotient $M$ of $\tilde M$ using the method of images.  For reasons explained below, we refer to this state as the Hartle-Hawking state $|0_{HH, M} \rangle$ on $M$.  In fact, since free fields $\phi$ on $M$ admit a ${\mathbb Z}_2$ symmetry $\phi \rightarrow - \phi$, we may in principle consider two such states  $|0_{HH, M} \rangle_\pm$ defined using either periodic or anti-periodic boundary conditions around the new homotopy cycle $\gamma$ in $M$.  Since we will concentrate on the periodic case $|0_{HH, M} \rangle_+$ below, we will also denote this state by $|0_{HH, M} \rangle$ with no subscript.

Since $M$ is globally hyperbolic, it contains no closed causal curves.  Thus the image $J\tilde x$ of any $\tilde x\in \tilde M$ never lies in either the causal future or past of $\tilde x$.  And since $M$ is smooth, $\tilde x$ and $J\tilde x$ cannot coincide.  Thus $\tilde x$ and $J\tilde x$ are spacelike related and quantum fields at $\tilde x$ commute with those at $J\tilde x$.  As a result, in linear quantum field theory, one may define quantum fields $\phi_\pm$ on $M$ in terms of quantum fields $\tilde \phi$ on $\tilde M$ via the relations
\begin{equation}
\label{eq:phidef}
\phi(x)_\pm = \frac{1}{\sqrt{2}}\left[\tilde \phi(\tilde x) \pm \tilde \phi (J\tilde x)\right],
\end{equation}
where $(\tilde x, J\tilde x)$ are the two points in $\tilde M$ that project to $x \in M$. Of course, in the antiperiodic ($-$) case, the overall sign of $\phi$ is not well-defined.  This case is best thought of as making $\phi$ charged under a ${\mathbb Z}_2$ gauge field with non-trivial holonomy around the ${\mathbb Z}_2$ cycle $\gamma$ of $M$.  Note that in either case $\phi(x)$ satisfies canonical commutation relations on a Cauchy slice of $M$ and so does indeed define a quantum field as claimed.

Any quantum state $\tilde \phi$ on $\tilde M$ then induces an associated quantum state $\phi_\pm$ on $M$.  In particular, this is true of the Hartle-Hawking state $|0_{HH,\tilde M}\rangle$, and we call the induced state $|0_{HH, M}\rangle_\pm$.  We will be interested in the expectation value in such states of the stress tensor  operator $T_{ab\pm}(x)$ (where the $\pm$ again refer to the choice of $\pm$ boundary conditions), and in particular the associated back-reaction on the spacetime $M$.  This back-reaction is most simply discussed by defining a new stress tensor $T_{ab\pm}(\tilde x)$ on $\tilde M$  as the pull-back of $T_{ab\pm}(x)$ under the natural projection $\tilde M \rightarrow M$.  In particular, for our Hartle-Hawking states we have
\begin{equation}
{}_\pm\langle 0_{HH, M}|
T_{ab\pm}(x)
|0_{HH, M}\rangle_\pm=
\langle 0_{HH, \tilde M}|T_{ab\pm}(\tilde x)
|0_{HH, \tilde M}\rangle.
\end{equation}
for any $\tilde x$ that projects to $x$.
The difference between $T_{ab}(\tilde x)$ and the stress tensor $\tilde T_{ab}(\tilde x)$ of the quantum field $\tilde \phi$ on $\tilde M$ will be made explicit below, but the important point is that the construction of the former involves the isometry $J$ which fails to commute with the Killing symmetry of $\tilde M$; see figure \ref{fig:geons}.  So while the expectation value of $\tilde T_{ab}(\tilde x)$ in the Hartle-Hawking state $|0_{HH, \tilde M}\rangle$ is invariant under the Killing symmetry, this property does {\it not} hold for the pull-back $T_{ab}(\tilde x)$ of $T_{ab}(x)$.

The point of pulling-back the stress tensor to $\tilde M$ is to reduce the analysis of back-reaction to calculations like that in \cite{Gao:2016bin}.  Since the (Hartle-Hawking) expectation value of
$T_{ab\pm}(\tilde x)$ is invariant under the action of $J$, the back-reaction of $T_{ab\pm}(x)$ on $M$ is just the ${\mathbb Z}_2$ quotient under $J$ of the back-reaction of $T_{ab\pm}(\tilde x)$ on $\tilde M$.   Since $\tilde M$ has a bifurcate horizon, after back-reaction traversability of the associated wormhole is related to the integral of $T_{ab\pm}(\tilde x)k^a k^b$ over the null generators of the horizon.   In particular, with sufficient symmetry (as in section \ref{sec:static}) the wormhole is traversable if and only if this value is negative along some generator.  More generally,  the wormhole can become traversable  only if this integral is negative along some generator \cite{Friedman:1993ty,Galloway:1999bp} and, as we will discuss in section \ref{sec:KKZBO} below, in our contexts traversability will be guaranteed if the average of this integral over all generators is negative.

To allow explicit formulae, we now specialize to the case of scalar fields.
The stress tensor of a free scalar field of mass $m$ takes the form
\begin{equation}
T_{ab\pm} = \partial_a \phi_\pm \partial_b \phi_\pm - \frac{1}{2} g_{ab} g^{cd} \partial_{c}\phi_\pm \partial_{d}\phi_\pm - \frac{1}{2}g_{ab}m^2\phi^2_{\pm}.
\end{equation}
In general, this diverges and requires careful definition via regularization (e.g., point-splitting) and renormalization.  However, using \eqref{eq:phidef}, the symmetry under $J$ of the actual stress energy $\tilde T_{ab}(\tilde x)$ of the quantum field $\tilde \phi$ on $\tilde M$, and the fact that $k^a$ is null we find
\begin{equation}
\label{eq:expand}
{}_\pm\langle 0_{HH,  M}|
T_{ab\pm}k^ak^b(x)
|0_{HH, M}\rangle_\pm
 =
\langle 0_{HH,\tilde M}|
\left[\tilde T_{ab}k^ak^b(\tilde x)
\pm  k^a k^b \partial_a \phi(\tilde x) \partial_b \phi(J\tilde x)\right]
| 0_{HH,\tilde M}\rangle.
\end{equation}
The second term on the right in \eqref{eq:expand} is manifestly finite since $\tilde x, J\tilde x$ are spacelike separated (and would be so even without contracting with $k^ak^b$).  Renormalization of $T_{ab\pm}$ is thus equivalent to renormalization of the stress tensor $\tilde T_{ab}$ of the $\tilde \phi$ quantum field theory on the covering space $\tilde M$.  However, when evaluated on the horizon and contracted with $k^ak^b$, any smooth symmetric tensor $Q_{ab}$ on $\tilde M$ that is invariant under the Killing symmetry must vanish\footnote{This is most easily seen by the standard argument that if $\xi^a$ is the Killing vector field then $Q_{ab}\xi^a\xi^b$ is smooth scalar invariant under the symmetry.  It is thus constant along the entire bifurcate horizon, and so must vanish there since $\xi^a$ vanishes on the bifurcation surface.  But $Q_{ab}k^a k^b \propto Q_{ab}\xi^a\xi^b$ on the horizon away from the bifurcation surface, so it must vanish there as well.  Smoothness then also requires $Q_{ab}\xi^a\xi^b$ to vanish on the bifurcation surface. This comment also justifies our use of Einstein-Hilbert gravity, as the first-order perturbative contributions from any higher derivative terms will vanish for the same reason.}.  As a result, the divergent terms in $\tilde T_{ab}k^ak^b(\tilde x)$ (which are each separately smooth geometric tensors with divergent coefficients) vanish on the horizon in all states, and invariance of the Hartle-Hawking state $| 0_{HH,\tilde M}\rangle$ means that the finite part of $\tilde T_{ab}k^ak^b(\tilde x)$ also gives no contribution to \eqref{eq:expand}.   Thus we have
\begin{equation}
\label{eq:crossterm}
{}_\pm\langle 0_{HH,  M}|
T_{ab\pm}k^ak^b(x)
|0_{HH, M}\rangle_\pm
 =
\pm \langle 0_{HH,\tilde M}|
k^a k^b \partial_a \phi(\tilde x) \partial_b \phi(J\tilde x)
| 0_{HH,\tilde M}\rangle.
\end{equation}

This result shows the key point.  Unless the integral of the right-hand-side vanishes, it will be negative for some choice of boundary conditions $(\pm$).  With that choice, back-reaction will then render the wormhole traversable.  It thus remains only to study this integral in particular cases, both to show that it is non-zero and to quantify the degree to which the wormhole becomes traversable.  We perform this computation for the AdS$_3$ $\mathbb{RP}^2$ geon and a simple Kaluza-Klein end-of-the-world brane in section \ref{sec:static}, and for a related example involving Kaluza-Klein zero-brane orbifolds in section \ref{sec:KKZBO}.

\subsection{A path integral perspective}
\label{sec:PI}

Before proceeding to explicit calculations, this section takes a brief moment to provide some useful perspective on the above construction, the relation to AdS/CFT, and in particular the connection to recent work \cite{Maldacena:2018lmt} by Maldacena and Qi and the original traversable wormhole of Gao, Jafferis, and Wall \cite{Gao:2016bin}.  Readers focused on the detailed computations relevant to our examples may wish to proceed directly to sections \ref{sec:static} and \ref{sec:KKZBO} and save this discussion for a later time.

For the purposes of this section we assume that the Hartle-Hawking state $|0_{HH,\tilde M}\rangle$ on the covering space $\tilde M$ is given by a path integral over (half of) an appropriate Euclidean (or complex) manifold $\tilde M_E$ defined by Wick rotation of the Killing direction in $\tilde M$.  In rotating cases, this may also involve analytic continuation of the rotation parameter to imaginary values, or a suitable recipe for performing the path integral on a complex manifold\footnote{In the presence of super-radiance or instabilities this procedure gives a non-normalizeable state that is not appropriate for quantum field theory.  In such cases one often says that the Hartle-Hawking state does not exist \cite{Wald:1995yp}.}.  We further assume that (as in figure \ref{fig:geons}) the isometry $J$ maps the Killing field $\xi^a$ to $-\xi^a$.  Note that global hyperbolicity of $M$ requires $J$ to preserve the time-orientation of $\tilde M$ so that, since $J$ exchanges the right- and left-moving horizons, it is not possible for $J$ to leave $\xi^a$ invariant.

Following \cite{Louko:1998dj}, one can extend the isometry $J$ to act on the complexification $\tilde M_{\mathbb C}$ of $\tilde M$, and thus on the particular section $\tilde M_E$.  The quotient $M_E= \tilde M_E/J$ and the desired Lorentzian spacetime $M = \tilde M/J$ are then associated with the complex quotient $M_{\mathbb C} = \tilde M_{\mathbb C}/J $.  As a result, $M_E$ is an analytic continuation of $M$.

Furthermore, for free fields the path integral over (half of) $M_E$ defines a state that is related to $|0_{HH,\tilde M}\rangle$ via the method of images.  This state is thus $|0_{HH, M}\rangle_+$, and we may instead obtain $|0_{HH, M}\rangle_-$ by coupling the bulk theory to a background ${\mathbb Z}_2$-valued gauge field with non-trivial holonomy around the ${\mathbb Z}_2$ cycle associated with taking the quotient by $J$.  It is due to this direct Euclidean (or complex) path integral construction that we call $|0_{HH, M}\rangle_\pm$ Hartle-Hawking states.  Taking this as the definition, such Hartle-Hawking states on $M$ can also be introduced for interacting quantum fields.

Indeed, in the AdS/CFT context one can go even farther.  Let us suppose that $M_E$ is the dominant bulk saddle point of a gravitational path integral over asymptotically locally AdS (AlAdS) geometries with conformal boundary $\partial M_E$. Then following \cite{Maldacena:2001kr} the CFT state defined by cutting open the path integral on $\partial M_E$ (perhaps again coupled to a ${\mathbb Z}_2$ gauge field having non-trivial holonomy) is dual to our Hartle-Hawking state $|0_{HH,M}\rangle_\pm$ on the bulk manifold $M_E$ at all orders in the bulk semi-classical approximation.

\subsubsection{The zero temperature limit}

\begin{figure}[t]
        \begin{center}
        \begin{subfigure}{.49\linewidth}
        \centering
                \includegraphics[width=0.5\textwidth]{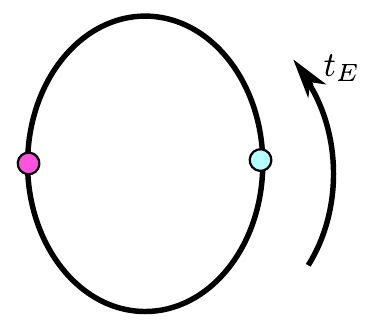}\quad
                \end{subfigure}
		\begin{subfigure}{.49\linewidth}
        \centering
                \includegraphics[width=0.35\textwidth]{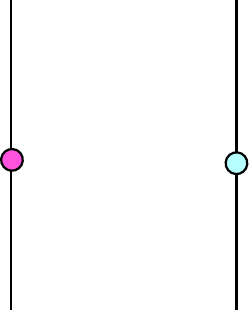}
                \end{subfigure}
        \end{center}
        \caption{ The boundary $\partial \tilde M_E$ of the non-extreme case (left) grows longer and longer to become  $\partial \tilde \Sigma \times {\mathbb R}$ in the extreme limit (right).}
        \label{fig:extremelim}
\end{figure}

Let us in particular consider the limit in which the temperature $T$ vanishes as defined by the Killing horizon in the bulk covering space $\tilde M$.  The Euclidean (or complex) period of $\tilde M_E$ diverges in this limit, so that $\tilde M_E$ can be approximated by $\tilde \Sigma \times {\mathbb R}$ for some manifold $\tilde \Sigma$ and $\partial \tilde M_E \rightarrow \partial \tilde \Sigma \times {\mathbb R}$; see figure \ref{fig:extremelim}.  Similarly, $M_E \rightarrow \Sigma \times {\mathbb R}$ and $\partial M_E \rightarrow \partial \Sigma \times {\mathbb R}$ for $\Sigma = \tilde \Sigma/J$.  So in the AdS/CFT context, we are studying the ground state of the CFT on $\partial \Sigma \times {\mathbb R}$.

This setting is now in direct parallel with that recently studied by Maldacena and Qi \cite{Maldacena:2018lmt}, which considered two copies of the SYK theory \cite{1993PhRvL..70.3339S,Kitaev} coupled through some multi-trace interaction and the associated two-boundary AdS$_2$ bulk dual to the Schwarzian sector of the SYK theory \cite{Maldacena:2016upp}.  From the CFT perspective, the multi-trace coupling is clearly critical to allow the two SYK models to interact.  From the bulk perspective, this coupling is again critical in allowing traversability, as without it the system would be invariant under separate time-translations along each of the two boundaries (associated with separate time-translations in each of the two SYK models). Preserving this symmetry would then forbid any bulk solution in which the two boundaries are connected.  In our setting, there is generally just a single time-translation symmetry of $\Sigma \times {\mathbb R}$ along ${\mathbb R}$.

The formulation in terms of ground states was useful in the non-perturbative SYK analysis of \cite{Maldacena:2018lmt}.  It also provides a useful perspective on our perturbative bulk analysis.
In particular, since the bulk ground state will be invariant under Euclidean time-translations (see footnote \ref{foot:static}), any zero temperature wormholes must either be traversable at all times or not at all.   Now, noting that a trip through a traversable wormhole can be started at arbitrarily early times, but that (unless the wormhole is eternal) there is generally a latest time $t_f$ at which such a trip may be begun, we can use $t_f$ to quantify the extent to which a given wormhole is traversable\footnote{It is in fact more natural to use $t_f -t_i$, where $t_i$ is the earliest time at which a {\it past}-directed causal curve can traverse the wormhole.  But we implicitly assume some symmetry that includes time-reversal (e.g., $(t,\phi) \rightarrow (-t,-\phi)$) in the main text.}. So if the finite $T$ wormholes become traversable, and if perturbative calculations indicate that $t_f$ increases as $T \rightarrow 0$, then we may take this as an indication that the wormhole is both traversable and static (eternal) in the actual bulk ground state.  Consistent with \cite{Maldacena:2018lmt}, we will find indications in section \ref{sec:KKZBO} that this occurs in the presence of sufficiently many bulk fields.

\begin{figure}[t]
        \begin{center}
        \begin{subfigure}{.49\linewidth}
        \centering
         \includegraphics[width=0.80\textwidth]{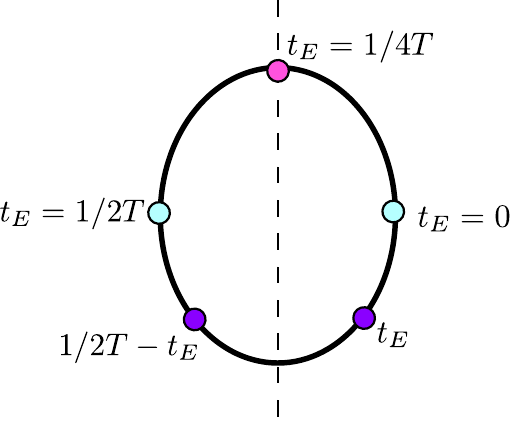}
        \end{subfigure}
        \begin{subfigure}{.49\linewidth}
        \centering
        \includegraphics[width=0.35\textwidth]{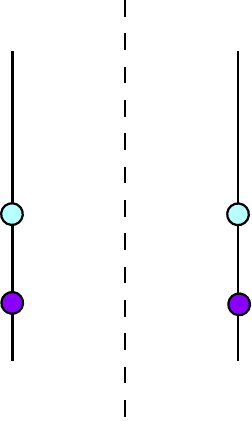}
        \end{subfigure}
        \end{center}
        \caption{The ground states of Maldacena and Qi \cite{Maldacena:2018lmt} can be obtained as limits of path integrals dominated by finite-temperature backgrounds in the bulk semi-classical limit.  At finite $T$, these path integrals would yield thermo-field double states if not for the additional multi-trace interaction that is bi-local in the Euclidean time $t_E$, coupling points $t_E$ and $\frac{1}{2T} - t_E$ related by reflection about the vertical dashed line.}
        \label{fig:MQPI}
\end{figure}

As a final comment, even if one is most interested in $T=0$, we see that the finite temperature setting is useful for performing perturbative computations.  A corresponding finite-$T$ version of \cite{Maldacena:2018lmt} can be obtained by studying SYK on a thermal circle defined by periodic Euclidean time $t_E$, so that slicing the circle at both $t_E=0$ and the antipodal point $t_E = 1/(2T)$ yields two-copies of SYK.  Introducing a multi-trace interaction that is non-local in $t_E$, and which in particular couples $t_E=0$ with $t_E = 1/(2T)$, then reproduces the ground state path integral of \cite{Maldacena:2018lmt} in the limit $T \rightarrow 0$ so long as one focuses on Euclidean times $t_E$ near both $t_E=0$ and $t_E = 1/(2T)$ and takes the non-local coupling to become time-independent in these regions.  For example, the coupling might take the form $g_T(Tt_E){\cal O}(t_E){\cal O}(\frac{1}{2T}-t_E)$ where $g_T$ is symmetric under $Tt_E\rightarrow \frac{1}{2} -Tt_E$; see figure \ref{fig:MQPI}. In field-theoretic cases (as opposed to the 0+1 SYK context), one may also wish to require that $g$ vanish at $t_E=\pm\frac{1}{4T}$ to prevent additional UV singularities.  At finite temperature, the Euclidean time-translation invariance is then broken by this non-local coupling, just as it is broken in our setting by the ${\mathbb Z}_2$ quotient of $\tilde M_E$ by $J$.   We also note that Wick rotation to Lorentz signature and appropriate choice of the resulting real-time coupling $g(t)$ then gives essentially the original traversable wormhole setting of \cite{Gao:2016bin}, though with Feynman boundary conditions instead of the retarded boundary conditions used in \cite{Gao:2016bin}.

\section{Simple traversable AdS$_3$ wormholes from Hartle-Hawking states}
\label{sec:joint}
\label{sec:static}
The non-rotating AdS$_3$ ${\mathbb{R P}}^2$ geon and KKEOW brane that form our first examples were defined in figure \ref{fig:geons} as simple $\mathbb{Z}_2$ quotients of BTZ and BTZ $\times S^1$ under appropriate isometries $J$.  Since quantum fields on the latter can be Kaluza-Klein reduced to an infinite tower of quantum fields on BTZ, it is clear from section \ref{sec:general} that both cases may be studied by computing the right-hand-side of \eqref{eq:crossterm} as defined by the two-point function of a single scalar field in the BTZ Hartle-Hawking state.

\subsection{BTZ and back-reaction}

As is well known, BTZ is itself a quotient of AdS$_3$, and the BTZ Hartle-Hawking two-point function is induced\footnote{Due to the fact that AdS$_3$ is an infinite cover of BTZ, this construction is slightly different than that discussed in section \ref{sec:free}.} via the method of images with periodic boundary conditions from the corresponding two-point function in the AdS$_3$ vacuum $|0\rangle_{\rm{AdS}_3}$. Since the latter is available in closed form, this construction provides a useful starting point for detailed calculations.

At this stage it is useful to introduce Kruskal-like coordinates $(U,V,\phi)$ on (non-rotating) BTZ.  We choose them so that the BTZ metric is
\begin{equation}
\label{eq:StaticBTZ}
\text{d}s^2 = g_{ab} \text{d}x^a \text{d}x^b =\frac{1}{(1+ UV)^2} (-4 \ell^{2} \text{d}U \text{d}V + r_+^2 (1-U V)^2 \text{d}\phi^2)
\end{equation}
where $\phi$ is periodic with period $2\pi$.  Such coordinates in particular allow us to write explicit expressions for the isometries $J$. For the AdS$_3$ $\mathbb{RP}^2$ geon, we take $J_{\rm geon}(U,V,\phi) = (V,U, \phi+\pi)$; i.e., it is given by reflecting the conformal diagram \ref{fig:geons} (right) about the dashed vertical line and acting with the antipodal map on the BTZ $\phi$-circle.  For the KKEOW brane, there is an additional periodic angle $\theta \in [0, 2\pi)$ on the internal $S^1$ and we take $J_{\rm eow}(U,V,\phi,\theta) = (V,U,\phi,\theta+\pi)$; i.e., this action is similar to $J_{\rm geon}$ but with the antipodal map acting on the internal $S^1$ as opposed to the BTZ $\phi$-circle.

As discussed in section \ref{sec:general}, the integral $\int \text{d}\lambda \langle T_{ab}\rangle k^a k^b$ along horizon generators will play a primary role in our analysis.  Here $\lambda$ is an affine parameter and $k^a$ the associated tangent vector.  In particular, since $U$ is an affine parameter along the BTZ horizon $V=0$, it will be useful to take $\lambda = U$ and $k^a\partial_a = \frac{\partial}{\partial U}$.

Let us begin with the observation that (as in \cite{Gao:2016bin}), at linear order the geodesic equation implies a null ray starting from the right boundary in the far past to have
\begin{equation}
\label{eq:geodesic}
V(U) = -(2 g_{UV}(V=0))^{-1}\int_{-\infty}^{U} \text{d}U h_{kk}
\end{equation}
where $h_{kk}$ is the norm of $k^a$ after first-order back-reaction from the quantum stress tensor (since $g_{ab}k^ak^b=0$) and we have used the fact that the background metric \eqref{eq:StaticBTZ} has constant $g_{UV}$ along the horizon $(V=0)$.

It thus remains to integrate $h_{kk}$. Since our $\mathbb{RP}^2$ geon and KKEOW brane both preserve rotational symmetry, this integral can be performed following \cite{Gao:2016bin}.
Defining $T_{kk} := T_{ab} k^a k^b$, the linearized Einstein equations give
\begin{equation}\label{eq:difeqh}
\begin{aligned}
\frac{1}{2}\left[\ell^{-2}(h_{kk} + \partial_U(U h_{kk})) - r_+^{-2}\partial_U^2h_{\phi\phi}\right] = 8\pi G_N \langle T_{kk}\rangle.
\end{aligned}
\end{equation}
To find the shift $\Delta V$ at $U=+\infty$, one merely integrates this equation over all $U$ to find
\begin{equation}
8\pi G_N  \int \text{d} U \langle T_{kk}\rangle = \frac{1}{2}\ell^{-2} \int \text{d}U h_{kk},
\end{equation}
where we have used asymptotically AdS boundary conditions and the requirement that the boundary stress tensor be unchanged at this order to drop the additional boundary terms\footnote{In the presence of scalars with $\Delta < 1$ (see below), the metric can receive large corrections near the boundary. But in AdS$_3$ such corrections give only a conformal rescaling of the original metric and so cannot contribute to \eqref{eq:geodesic}.  The specification that the boundary stress tensor be unchanged determines the choice of boundary gravitons -- or in other words the choice of linearized diffeomorphism along with the change in gravitational flux threading the wormhole -- to be added along with the perturbation.}.
 Thus,
\begin{equation}
\label{eq:DeltaVfinal}
\Delta V(+\infty) = -\frac{8 \pi G_N \ell^2}{g_{UV}(0)} \int_{-\infty}^{\infty}\text{d}U \langle T_{kk}\rangle = 4 \pi G_N  \int_{-\infty}^{\infty}\text{d}U \langle T_{kk}\rangle.
\end{equation}

Similarly, if we are interested in measuring the shift at the center of the wormhole ($U=V=0$), we can integrate equation \eqref{eq:difeqh} from $U=-\infty$ to $U=0$. The contribution from $\partial_U (U h_{kk})$ again vanishes, as $U h_{kk}|_{U=0}=0$. We thus find
\begin{equation}
\Delta V(0) = -\frac{8 \pi G_N \ell^2}{g_{UV}(0)} \int_{-\infty}^{0}\text{d}U \langle T_{kk}\rangle = 2\pi G_N  \int_{-\infty}^{\infty}\text{d}U \langle T_{kk}\rangle =  \frac{1}{2}\Delta V(+\infty),
\end{equation}
where we have used the fact that in our examples $\langle T_{kk}\rangle$ is also symmetric about $t=0$.  This quantity gives a measure of the length of time that the wormhole remains open as measured by an observer at the bifurcation surface.  Since the result is simply related to the shift at the left boundary, it will be convenient below to define $\Delta V : = \Delta V(+\infty)$ and to understand that all quantities of interest are simply related to this $\Delta V$.

For example, we might also like to compute the minimum length of time it takes to travel through the wormhole. Note that at first order in perturbation theory, any null ray that traverses the wormhole (from right to left) will be perturbatively close to $V=0$.  As a result, at this order it will differ from \eqref{eq:geodesic} by at most a constant off-set; i.e.,
\begin{equation}
\label{eq:gengeo}
\Delta V(U) := V(U) - V(-\infty)= -\int_{-\infty}^{U}\text{d}U \frac{
h_{kk}}{2 g_{UV}(V=0)}.
\end{equation}
Choosing a conformal frame in which the boundary metric is $\text{d}s^2_{\partial \rm BTZ} = -\text{d}t^2 +\ell^2 \text{d}\phi^2$, we find on the boundary
$\text{d}t^2 = \frac{\ell^4\text{d}V^2}{r_+^2V^2}$, so we may
choose $t =\pm\frac{\ell^2}{2r_+}\ln\left( \pm\frac{V}{\ell}\right)$, with the choice of signs $(\pm)$ being both $(+)$ on the right boundary and both $(-)$ on the left. Since the wormhole is traversable for $\Delta V<0$, the shortest transit time $t_*$ from the right to left boundary is realized by the geodesic that leaves the right boundary at $V = -\Delta V/2$ and arrives at the left boundary at $V=\Delta V/2$. We
 thus find
\begin{equation}
t_* = -\frac{\ell^2}{r_+}\ln\left( \frac{|\Delta V|}{2\ell}\right).
\end{equation}

\subsection{Ingredients for the stress tensor}

The quotient of AdS$_3$ used to obtain BTZ is associated with the periodicity of $\phi$.  As a result, taking $\phi$ in \eqref{eq:StaticBTZ} to range over $(-\infty, \infty)$ yields a metric on a region of empty global AdS$_3$.

Now, at spacelike separations (as appropriate for $\tilde x, J\tilde x$), the AdS$_3$  two-point function for a free scalar field of mass $m$ is determined by its so-called conformal weight
\begin{equation}
\Delta = 1 \pm \sqrt{1+m^2 \ell^2},
\end{equation}
where the choice of $\pm$ is associated with a choice of boundary conditions, though for $m^2 \ge 0$ only the (+) choice is free of ghosts \cite{Andrade:2011dg}. The AdS$_3$ two-point function is then (see section 4.1 of reference \cite{Ichinose:1994rg})
\begin{equation}
\label{eq:AdS2pt}
G(x,x') = G_{\rm{AdS}_3}(Z) = \frac{1}{4 \pi} (Z^2-1)^{-1/2}(Z+ (Z^2-1)^{1/2})^{1-\Delta}
\end{equation}
where $Z = 1+\sigma(x,x')$ and $\sigma(x,x')$ is half of the (squared) distance between $x$ and $x'$ in the four dimensional embedding space\footnote{In reference \cite{Louko:2000tp}, this distance was called the ``chordal distance'' in the embedding space. Here, $\sigma(x,x')$ is half of this chordal distance.}, and with all fractional powers of positive real numbers defined by using the positive real branch.  The BTZ two-point function is
\begin{equation}
\label{eq:GBTZ}
G_{\rm{BTZ}}(\tilde x,\tilde x') = \frac{1}{4 \pi} \sum_{n \in {\mathbb Z}} (Z_n^2-1)^{-1/2}(Z_n+ (Z_n^2-1)^{1/2})^{1-\Delta},
\end{equation}
where $Z_n = 1+\sigma(x,x'{}_n)$ where $x$ is any point in AdS$_3$ that projects to $\tilde x$ in BTZ and $x'{}_n$ are the inverse images in AdS$_3$ of $\tilde x'$ in BTZ.
A standard calculation then gives
\begin{equation}
\begin{aligned}
\sigma \left( x,x'_n \right)=&\frac{\ell^2}{{(U V+1) (U' V'+1)}}\left[ (U V-1) (U'V'-1) \cosh \left(r_+ \left(\phi -\phi '_n\right)\right)\right.\\
&\left. -(U V+1) (U' V'+1)+2 (U V'+V U') \right]
\label{eq:tpf}
\end{aligned}
\end{equation}
in terms of our Kruskal-like BTZ coordinates.  Here we take $x = (U,V,\phi)$ (in either the geon/KKEOW brane or AdS$_3$) and $x'_n = (U',V',\phi'_n)$. As noted above, the $x'{}_n$ are related by $2\pi$ shifts of the BTZ $\phi$ coordinate so that $\phi'_n := \phi(x'{}_n) = \phi' + 2\pi n$ for some $\phi'$.

In computing \eqref{eq:crossterm}, we will set $\tilde x'=J\tilde x$ and thus $U'= V, V'=U$.  For the AdS$_3$ geon we also set $\phi' = \phi + \pi$, while $\phi'=\phi$ for our KKEOW brane.  So for each $n$ both cases involve computations of \eqref{eq:crossterm} that differ only by an overall shift of $\phi'$ by $\pi$.

\begin{figure}[t]
        \begin{center}
                \includegraphics[width=0.4\textwidth]{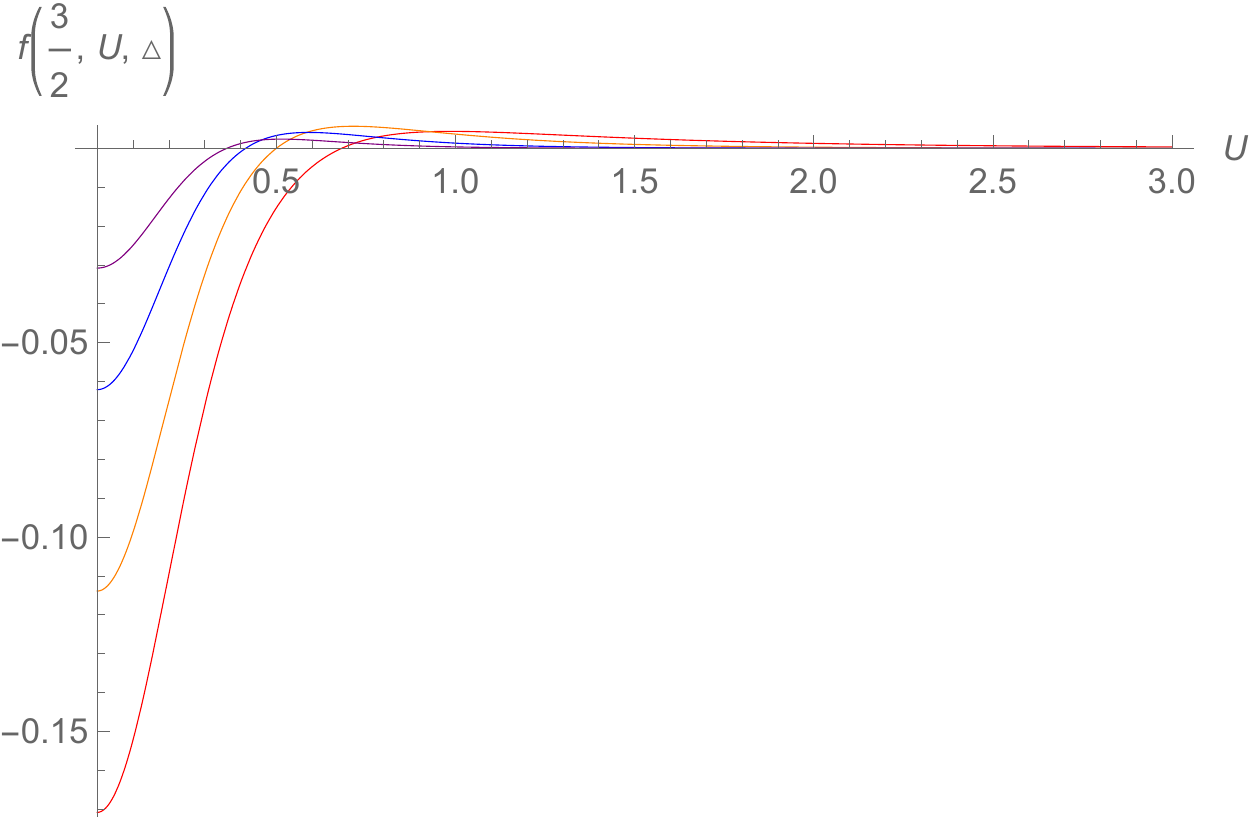}
                \includegraphics[width=0.4\textwidth]{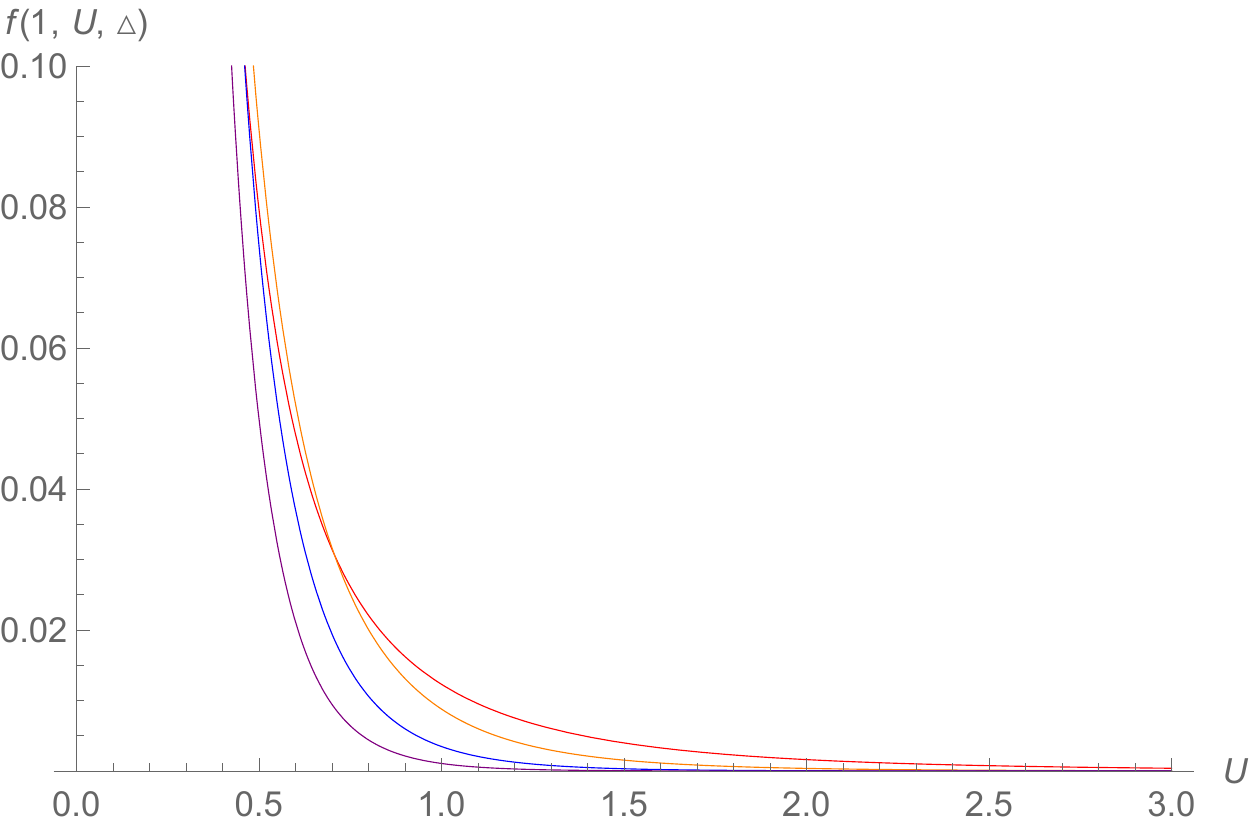}
                \end{center}
        \caption{Some of the functions \eqref{eq:f(C)}. {\bf Left:} $C=1.5$, for $\Delta =1$ (red), $\Delta =2$ (orange), $\Delta =3$ (blue), and $\Delta =4$ (purple). {\bf Right:} $C=1$, for $\Delta =1$ (red), $\Delta =2$ (orange), $\Delta =3$ (blue), and $\Delta =4$ (purple).}
        \label{fig:f}
\end{figure}

In fact, one sees immediately from \eqref{eq:tpf} that the integral of \eqref{eq:crossterm} depends only on $C \equiv \cosh\left(r_+(\phi-\phi'_n)\right)$.  For the geon case, this is $-2\pi (n_{\rm geon}+\frac{1}{2}) r^{\rm geon}_+$, while for the KKEOW brane it is $-2\pi n_{\rm eow} r_+^{\rm eow}$.  So for $r_+^{\rm geon} = 2 r^{\rm eow}_+$ and $n_{\rm eow} = 2n_{\rm geon} +1$ (for odd $n_{\rm eow}$) or $r_+^{\rm geon} = 4 r_+^{\rm eow}$ and $n_{\rm eow} = 4n_{\rm geon} +2$  (for even $n_{\rm eow}$), the two computations involve precisely the same integral over generators of the BTZ horizon.  Below, we briefly comment on this integral for general $C$ and then use it to obtain the desired geon and KKEOW brane results.  In particular, working on the horizon $V=0$ we define
\begin{equation}
\label{eq:integrand}
f(C,U; \Delta):=
\langle 0_{HH,\rm{AdS}_3}|
\partial_U \phi(x) \partial_U \phi(x')
| 0_{HH,\rm{AdS}_3}\rangle|_{V=0}
\end{equation}
for $x,x'$ as above in AdS$_3$.  Using \eqref{eq:AdS2pt} and \eqref{eq:tpf} then gives
\begin{equation}
\begin{aligned}
f(C,U; \Delta)=&\frac{\left(\sqrt{B^2-1}+B\right)^{-\Delta } }{2 \pi  \left(B^2-1\right)^{5/2}} \Bigl\{\left(B^2-1\right)^2 (1-\Delta ) \left(-\frac{2 B^2 U^2 }{\left(B^2-1\right)^{3/2}}+\frac{ 2 U^2 + B}{\sqrt{B^2-1}}+1\right)\\
&+ \left[  \left(B^2-1\right) \left( 2(\Delta^2 - \Delta-1)U^2 -B\right)   +4 B \sqrt{B^2-1} (\Delta -1) U^2 +6 B^2  U^2 \right] \\
& \times \left(\sqrt{B^2-1}+B\right)\Bigr\},
\end{aligned}
\label{eq:f(C)}
\end{equation}
where $B(U) \equiv 2U^2+C $. To give the reader a feel for this complicated-looking function, we plot $f$ in  figure \ref{fig:f} below for various values of $C$, $\Delta$.

We can also consider simple, limiting cases of $f(C,U;\Delta)$. For instance, when $\Delta =0,1,2$, this becomes
\begin{eqnarray}
f(C,U;0) = f(C,U;2) &=& \frac{1-C^2+2 C U^2+8 U^4}{2 \pi  \left[\left(C+2 U^2\right)^2-1\right]^{5/2}},\\
f(C,U;1) &=& \frac{C-{{C}^{3}}+4{{U}^{2}}-2{{C}^{2}}{{U}^{2}}+4C{{U}^{4}}+8{{U}^{6}}}{2\pi {{\left[{{\left( C+2{{U}^{2}} \right)}^{2}-1} \right]}^{5/2}}}.
\end{eqnarray}

\begin{figure}[t]
        \begin{center}
                \includegraphics[width=0.4\textwidth]{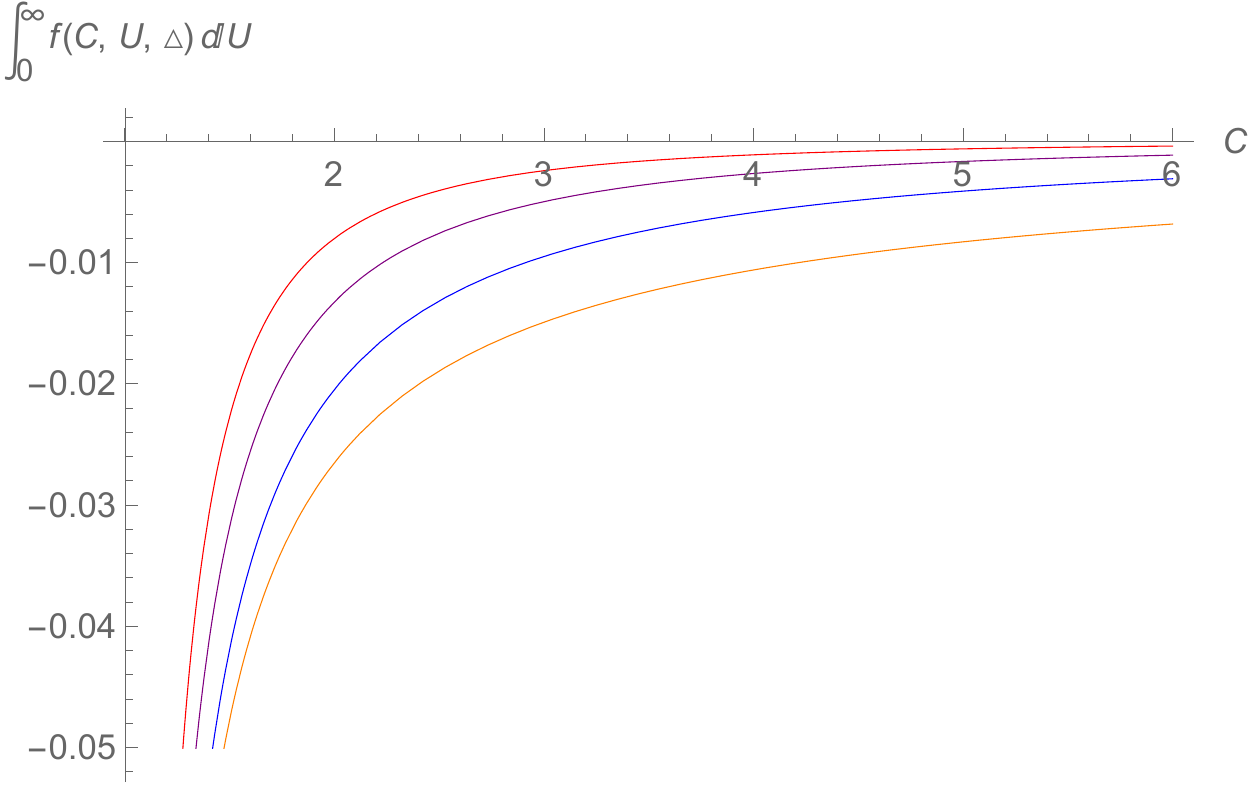}
                \includegraphics[width=0.4\textwidth]{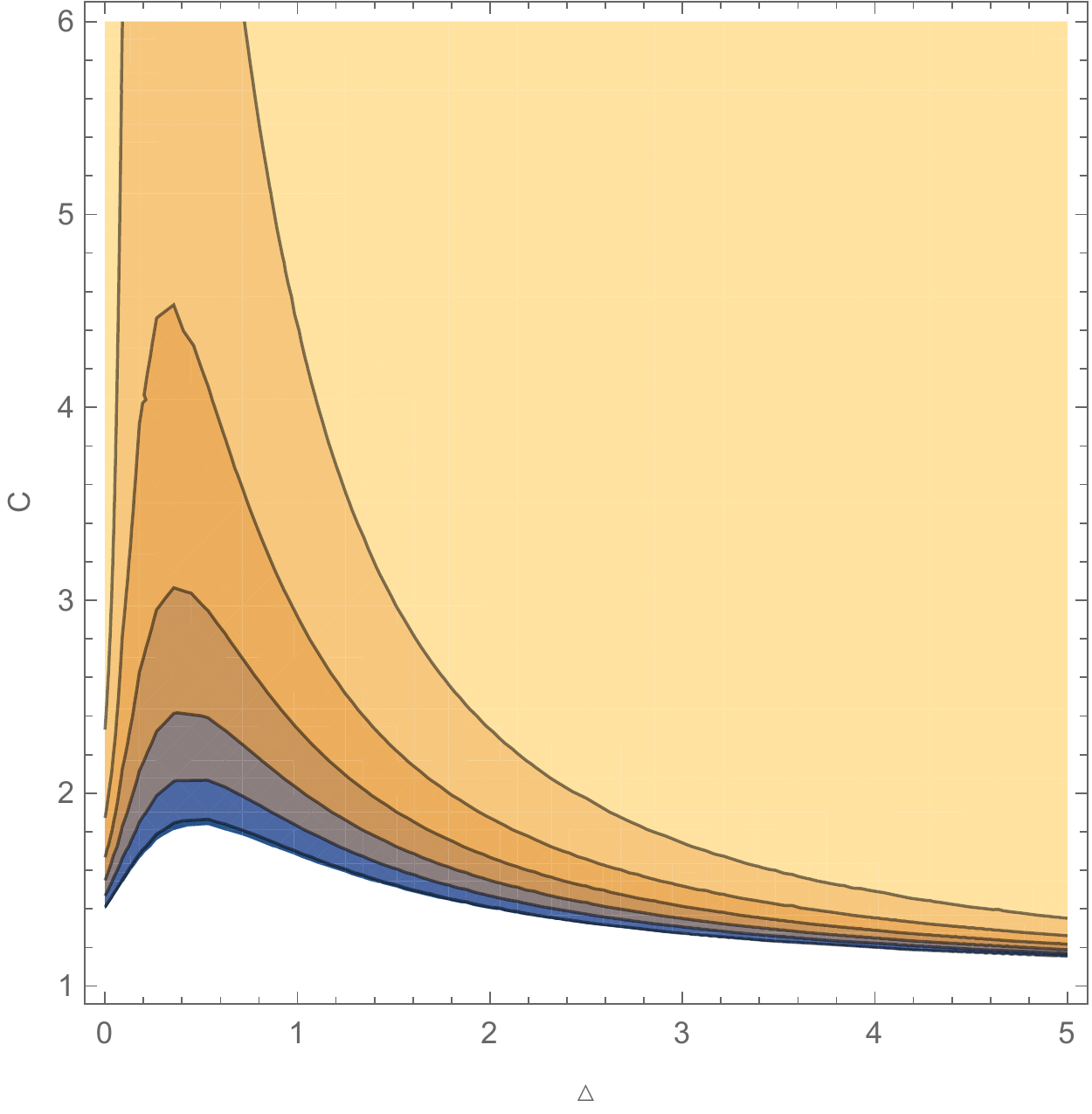}
                \includegraphics[width=0.09\textwidth]{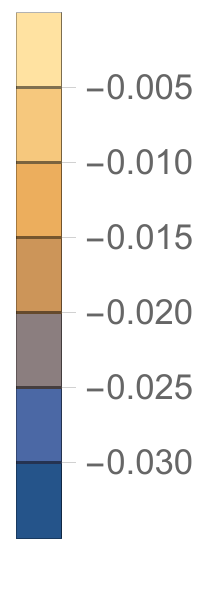}
                \end{center}
        \caption{{\bf Left:} $\int _0^\infty f(C,U;\Delta )\text{d}U$ as a function of $C$ for $\Delta =0$ (red), $\Delta =0.5$ (orange), $\Delta =1$ (blue), $\Delta =1.5$ (purple). {\bf Right:} Further numerical results for $\int _0^\infty f(C,U;\Delta )\text{d}U$ supporting \eqref{eq:conj}.  The white region is even more negative than those shown in color.}
        \label{fig:neg}
\end{figure}

More generally, when $\Delta = 0,1/2,1,3/2,2\dots $, the integral $\int \text{d}U f(C,U;\Delta)$ can be performed analytically. For example, we find
\begin{eqnarray}
\label{eq:Efunct}
\int_{0}^{\infty }{f(C,U;0)}\text{d}U=\int_{0}^{\infty }{f(C,U;2)}\text{d}U&=&\frac{\left( C-1 \right)K\left( \frac{2}{C+1} \right)-CE\left( \frac{2}{C+1} \right)}{8\sqrt{2}\pi \left( C-1 \right)\sqrt{C+1}} <0,\\
\int_{0}^{\infty }{f(C,U;1)}\text{d}U&=&-\frac{E\left( \frac{2}{C+1} \right)}{8\sqrt{2}\pi \left( C-1 \right)\sqrt{C+1}}<0,
\end{eqnarray}
for $C > 1$ where $K(k)$ is the complete elliptic integral of the first kind and $E(k)$ is the complete elliptic integral of the second kind.  Since both $E$ and $K$ are positive functions, the second inequality is manifest.  The first inequality can be seen from figure \ref{fig:neg} (left).

For other values of $\Delta$ and $C$, numerical integration suggests that the result continues to be negative as seen in figure \ref{fig:neg}:
\begin{equation}
\label{eq:conj}
\int_{0}^{\infty }f(C,U;\Delta) \text{d}U<0\text{, for }\Delta\ge 0, C>1.
\end{equation}
For all $\Delta \ge 0$, when $C\to 1^+$, the integral $\int_{0}^{\infty }f(C,U;\Delta) \text{d}U$ becomes divergent and goes to $-\infty $. In contrast, both $f$ and its integral vanish for all $\Delta$ as $C \rightarrow \infty$. For later use, we note that expression \eqref{eq:tpf} simplifies in the  $r_+\rightarrow 0$ limit, which gives
\begin{equation}
\label{eq:fC1}
\begin{aligned}
f(1,U;\Delta)=\frac{{{\left( U+\sqrt{1+{{U}^{2}}} \right)}^{2-2\Delta }}}{32\pi {{U}^{3}}{{\left( 1+{{U}^{2}} \right)}^{5/2}}}
& \left[ 1+2U\sqrt{1+{{U}^{2}}}\left( -1+\Delta  \right)+8{{U}^{3}}\sqrt{1+{{U}^{2}}}\left( -1+\Delta  \right) \right. \\
& \left. +4{{U}^{4}}\left( 2-2\Delta +{{\Delta }^{2}} \right)+{{U}^{2}}\left( 6-8\Delta +4{{\Delta }^{2}} \right) \right].
\end{aligned}
\end{equation}
Some of these functions are plotted in figure \ref{fig:f} (right).

\subsection{Traversability of the AdS $\mathbb{RP}^2$ geon}
\label{sec:geon}
\begin{figure}
\centering
\begin{subfigure}{.475\linewidth}
\centering
\includegraphics[width=1\linewidth]{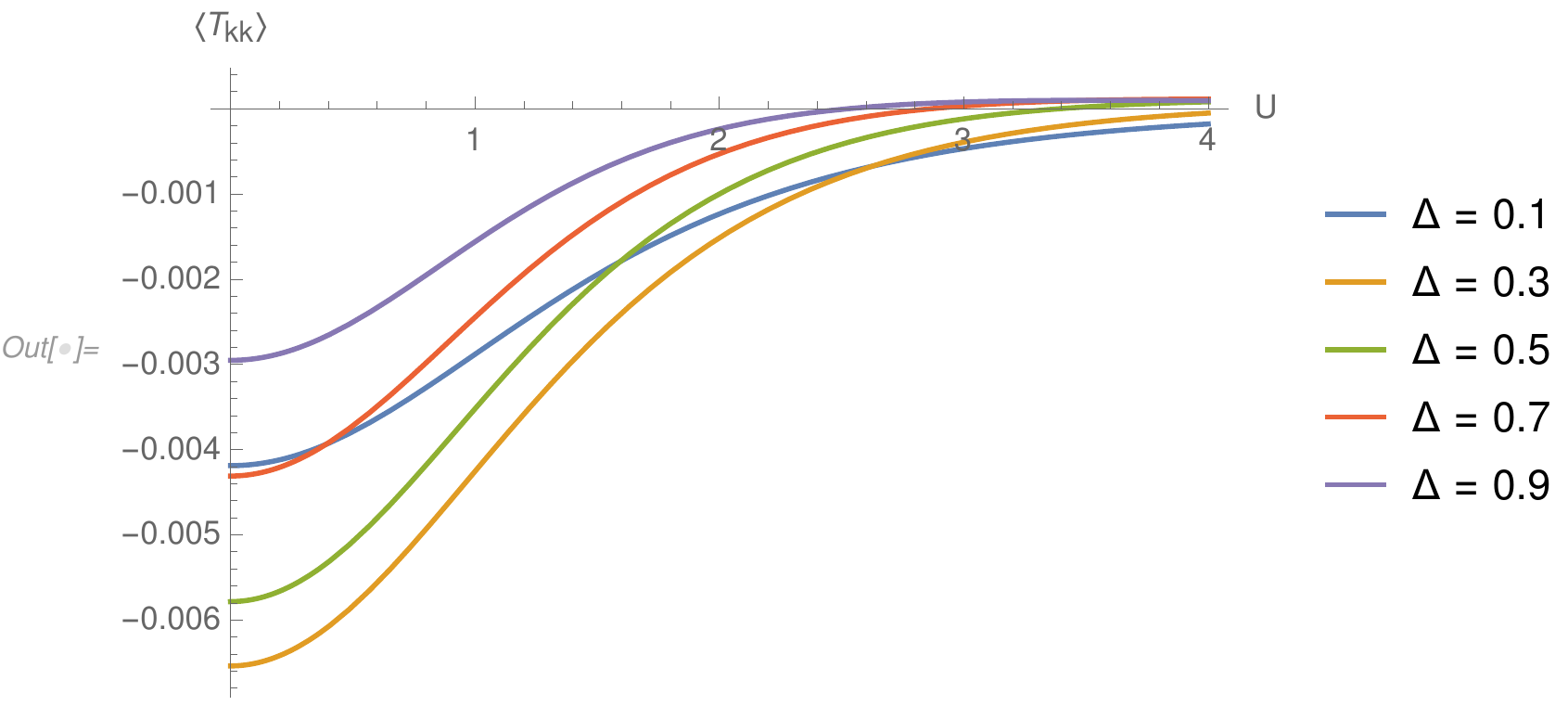}\quad%

\end{subfigure}
\begin{subfigure}{.475\linewidth}
\centering
\includegraphics[width=1\linewidth]{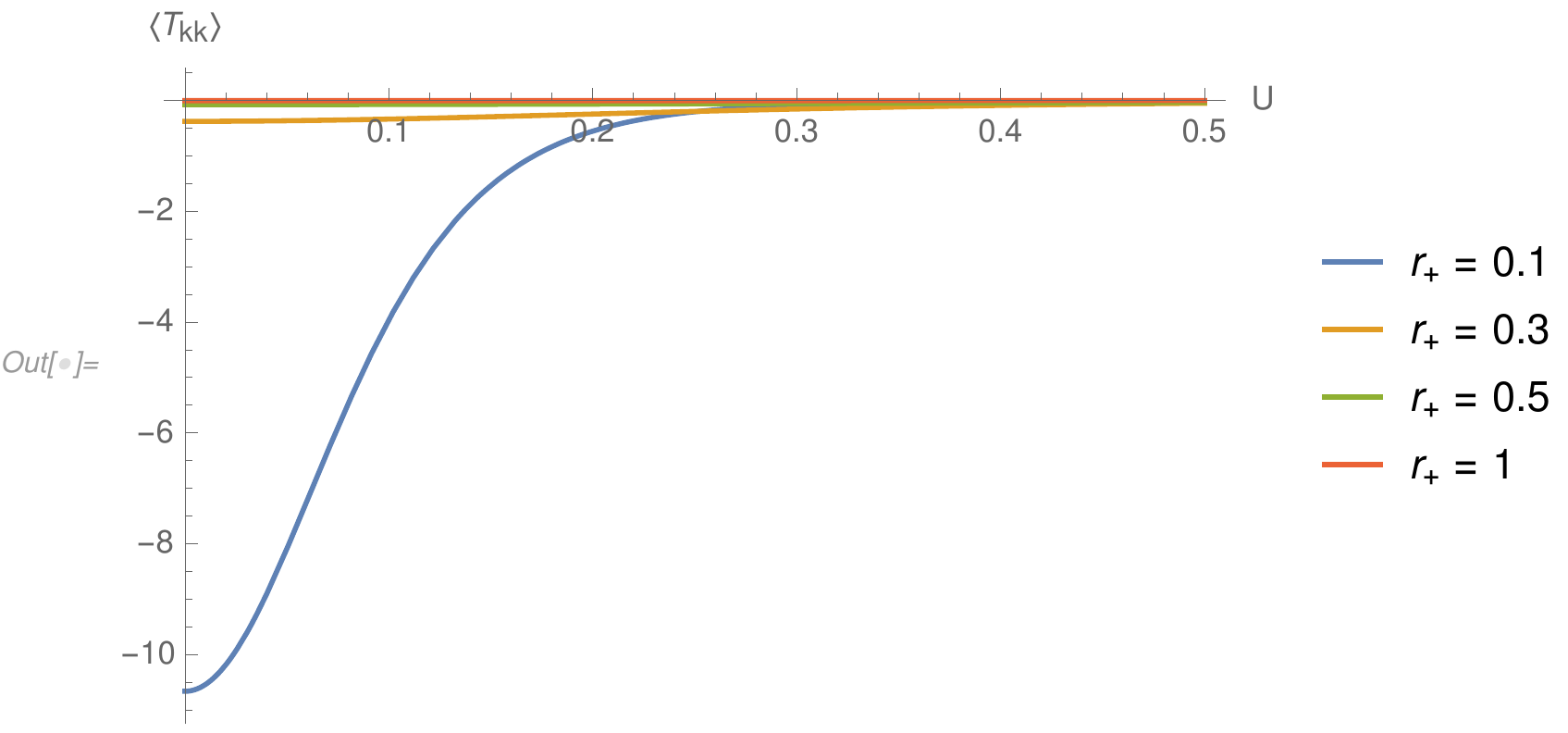}\quad%

\end{subfigure}
\caption{{\bf Left:} $\langle T_{kk+}\rangle$ on the AdS$_3$ geon for various $\Delta$ with $r_+ = \ell$. {\bf Right:} $\langle T_{kk+}\rangle$ on the AdS$_3$ geon for  various $r_+$ (in units of $\ell $) with $\Delta = 0.5$. }
\label{fig:nonrotatingtuu}
\end{figure}

\begin{figure}
\centering
\begin{subfigure}{.475\linewidth}
\centering
\includegraphics[width=1\linewidth]{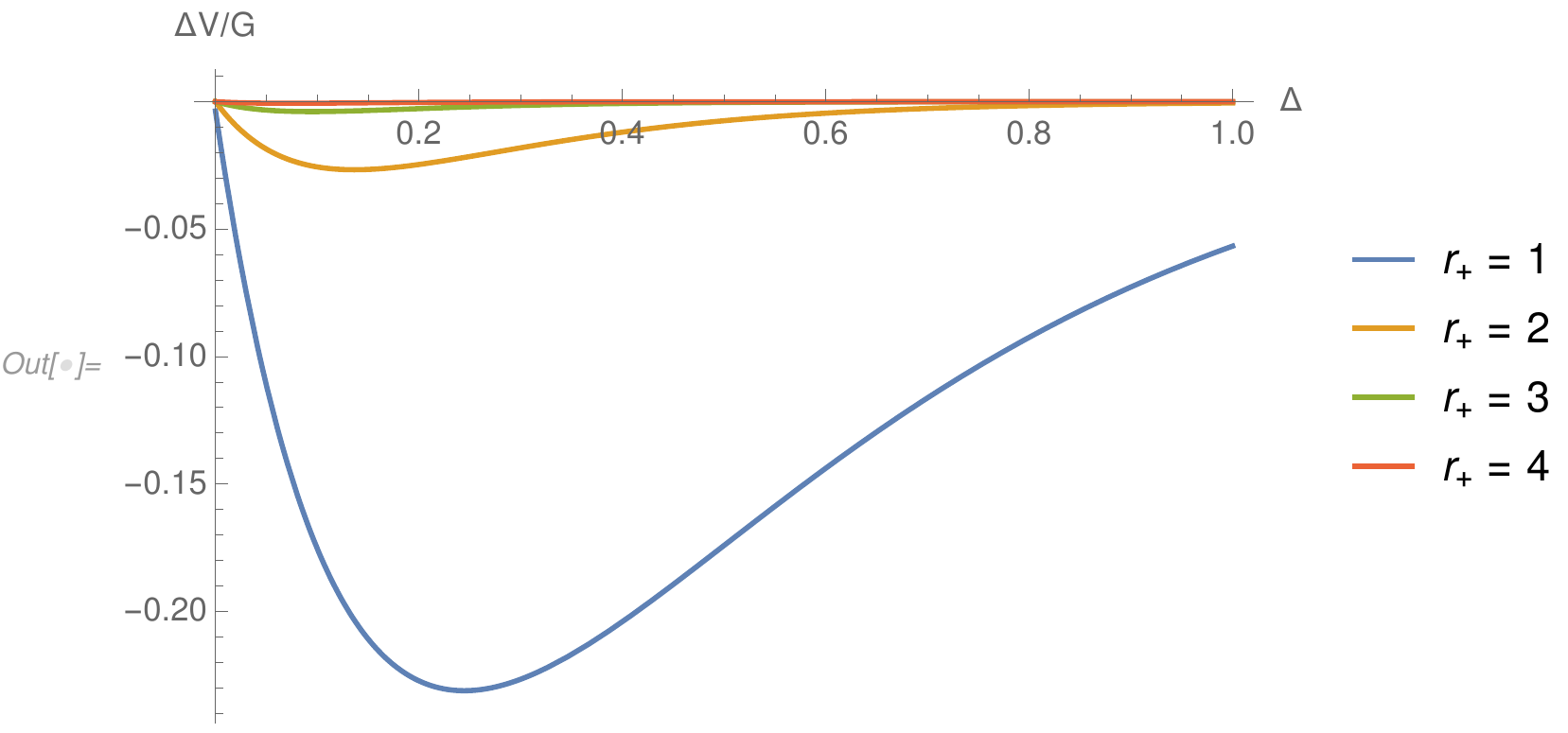}\quad%
\end{subfigure}
\begin{subfigure}{.475\linewidth}
\centering
\includegraphics[width=1\linewidth]{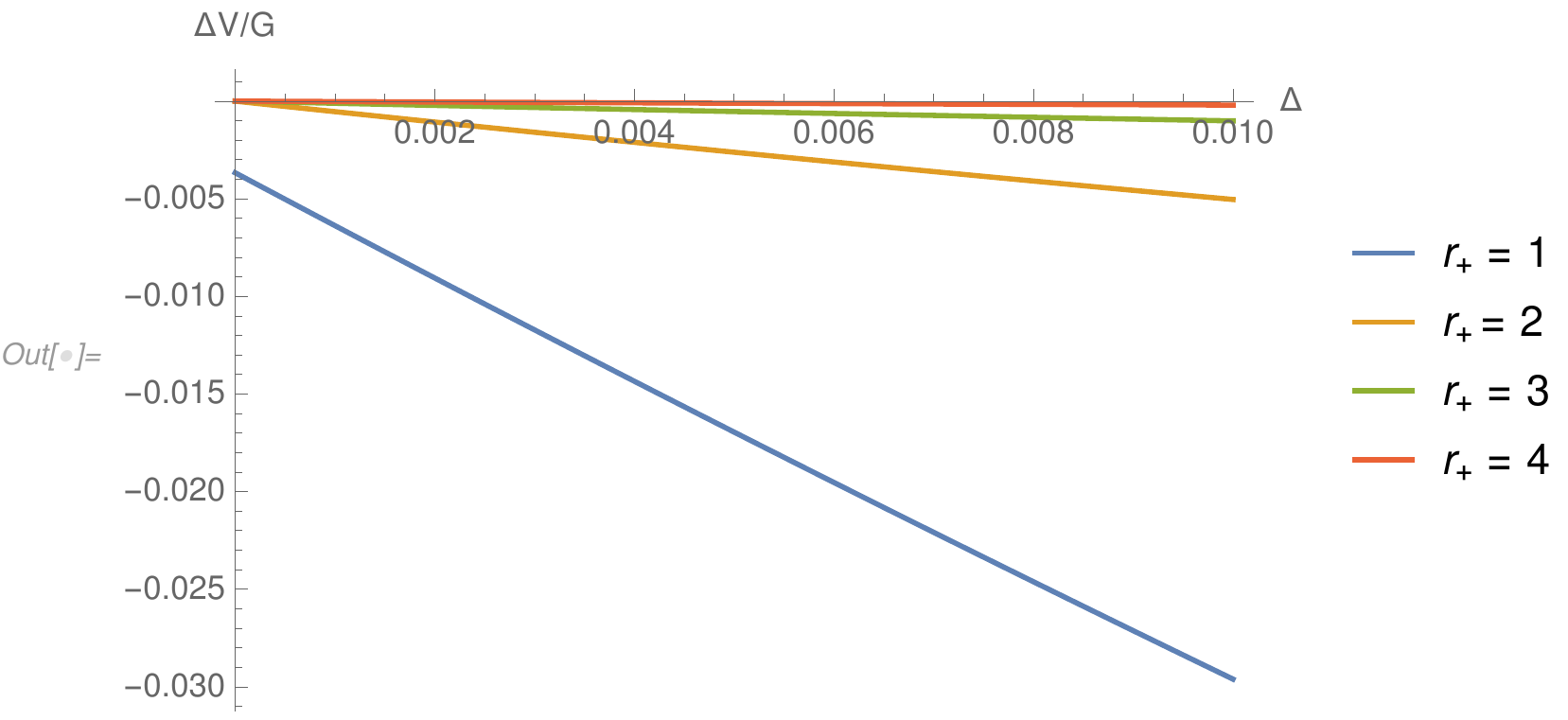}\quad%
\end{subfigure}
\caption{{\bf Left:} The shift $\Delta V$ as measured on the left boundary. {\bf Right:} $\Delta V$ remains negative at $\Delta =0$.}
\label{fig:nonrotatinginttuu}
\end{figure}

\begin{figure}
\centering
\includegraphics[width=0.4\linewidth]{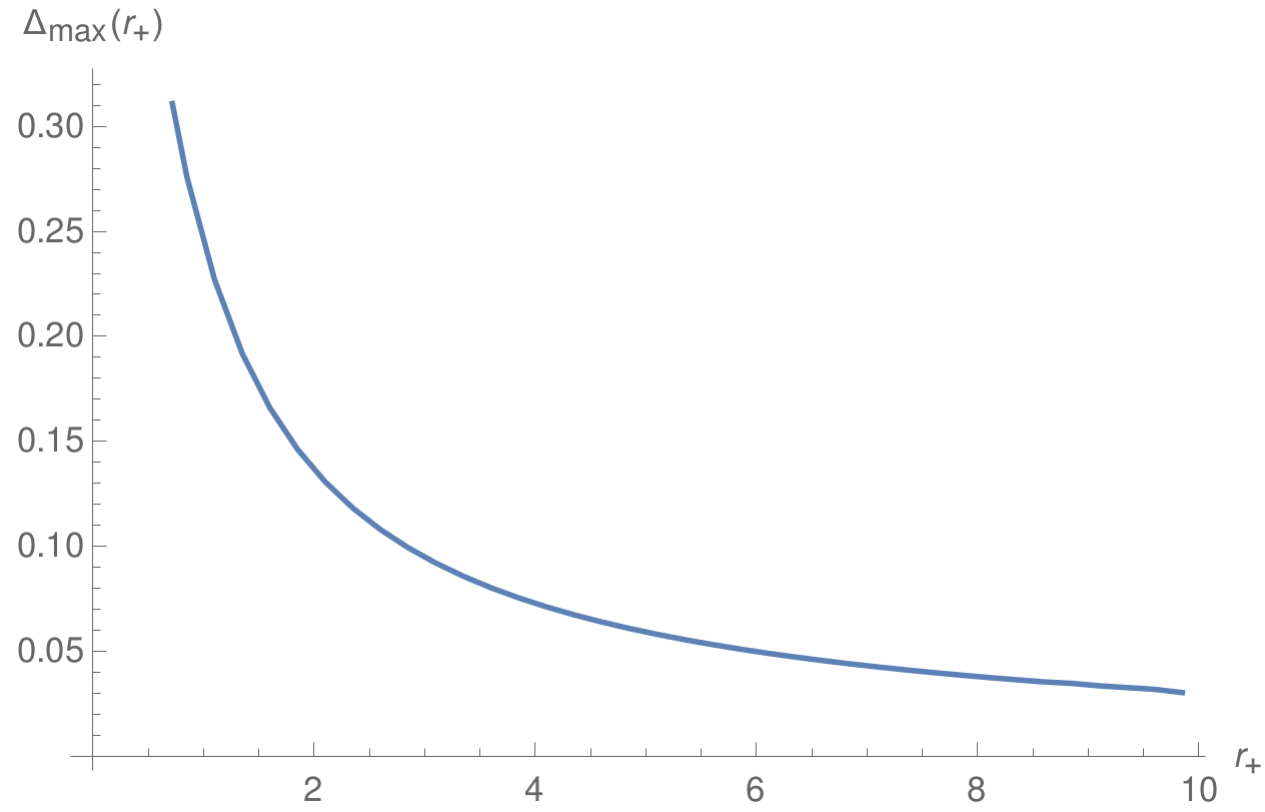}\quad%
\caption{The $\Delta$ which maximizes $|\Delta V|$ for a given $r_+$.}
\label{fig:DeltaMax}
\end{figure}

We can now use the above ingredients to study the traversability of the $\mathbb{RP}^2$ geon with back-reaction from a periodic $(+)$ scalar.  Since the analysis involves only a single bulk quantum field, we have
\begin{equation}
\label{eq:geonTkk}
\int \text{d}U \langle T_{kk+}\rangle = \sum_{n\in {\mathbb Z}} \int \text{d}U f(C_n,U;\Delta)
\end{equation}
for $C_n$ defined by  $\phi - \phi'_n = (2n+1)\pi$.  From \eqref{eq:conj} we can already see that the associated first-order back-reaction will make the wormhole traversable.  As pointed out in \cite{Gao:2016bin} and shown in figure \ref{fig:neg} (right), $\Delta V \rightarrow -\infty$ as $r_+ \rightarrow 0$.  But in contrast to \cite{Gao:2016bin}, it follows from \eqref{eq:Efunct} that $\Delta V$ remains finite as $\Delta \rightarrow 0$ (though it is numerically small, see figure \ref{fig:nonrotatinginttuu}).  Typical stress tensor profiles and horizon shifts $\Delta V$ are shown in figures \ref{fig:nonrotatingtuu} and \ref{fig:nonrotatinginttuu}, where we used Mathematica to numerically perform both the integral over $U$ and the sum over $n$ in \eqref{eq:geonTkk}. While the total stress energy is used in the figures, since $f$ decreases rapidly at large $C$, for $r_+ > 1$ there is little difference between $\langle T_{kk+}(U)\rangle$ and the $n=0$ term $f(C_0,U;\Delta)$ (except for a factor of $2$ that arises because $C_0=C_{-1}$ for the geon since these cases represent $\phi-\phi_n{}' = \pm \pi$).  An interesting feature of the results is that the value $\Delta_{\rm max}$ of $\Delta$ that maximizes $|\Delta V|$ depends strongly on $r_+$ as shown in figure \ref{fig:DeltaMax}.

\subsection{Traversability and the KKEOW brane}
\label{sec:EOTW}
Although it involves only a single scalar from the 4d perspective (say with 4d mass $m$), Kaluza-Klein reduction to $d=3$ gives a tower of scalar fields with effective 3d masses $m_{\text{eff}}\ell =\sqrt{m^2\ell ^2+\left(\ell /R_{S^1}\right)^2p^2}$ is the radius of the Kaluza-Klein circle. For each $p$, the corresponding effective conformal dimension $\Delta $ is then
\begin{equation}
\label{eq:Delta}
\Delta \left(k\right)=1\pm \sqrt{1+{{m}^{2}\ell ^2}+{{\left( \frac{{\ell}}{{{R}_{{{S}^{1}}}}} \right)}^{2}}{{p}^{2}}}.
\end{equation}
The choice of sign $(\pm)$ can be made independently for each $p$ so long as one allows boundary conditions that are non-local on the internal $S^1$.  But violating the CFT unitarity bound $\Delta >0$ leads to ghosts \cite{Andrade:2011dg}, so the (+) sign is required at large $p$.

Because each $p$ is associated with a wavefunction $e^{ip\theta}$ on the internal $S^1$, and since $J$ maps $\theta \rightarrow \theta + \pi$, the contribution to \eqref{eq:crossterm} from each $p$ is $(-1)^p$ times the result $\langle T_{kk}(U; \Delta) \rangle$ one would obtain from a single scalar of weight $\Delta$ on BTZ under the (singular) ${\mathbb Z}_2$ quotient by $(U,V,\phi) \rightarrow (V,U,\phi)$.  As a result, and using the symmetry under $p \rightarrow -p$ we have
\begin{equation}
\label{eq:nSum}
\left\langle {{T}_{kk+}}\left( U \right) \right\rangle =\sum\limits_{n\in \mathbb{Z}}{\hat f\left( {{C}_{n}},U;\ell/R_{S^1}\right)},
\end{equation}
where
\begin{equation}
\label{eq:pSum}
\hat f\left( {{C}_{n}},U;\ell/R_{S^1}\right)=f\left( {{C}_{n}},U;\Delta \left( 0 \right) \right)+2\sum\limits_{p=1}^{\infty }{{{\left( -1 \right)}^{p}}f\left( {{C}_{n}},U;\Delta \left( p \right) \right)}
\end{equation}
with $\phi - \phi'_n = 2\pi n$.

As discussed near \eqref{eq:fC1}, the function $f\left(C_n,U; \Delta \right)$ has a non-integrable singularity at $U=0$ for $n=0$ at each $\Delta$ but is finite for $n\ne 0$.  Yet since the full 4d spacetime is smooth, the 4d-stress tensor and the back-reacted metric must be smooth as well.  This occurs because the alternating signs in \eqref{eq:pSum} cause the $U=0$ singularities to cancel when summed over $p$.

For $U\neq 0$ the sums over $n$ and $p$ converge rapidly.  In particular, for each $n \neq 0$ the sum over $p$ converges exponentially since $f(C_n,U;\Delta(p))$ evaluates the BTZ two-point function at some fixed spacelike separation on BTZ set by $C_n$ for 3d fields that have large mass at large $p$. Indeed, for fixed $U\neq 0$ the same is true even when $n=0$. And the sum over $n$ is also exponentially convergent since $\sigma(\tilde x, J\tilde x)$ grows exponentially with $n$.  As a result, one approach to computing \eqref{eq:nSum} is to numerically perform the sums away from $U=0$ and then to recover the value at $U=0$ by taking a limit, though care will be required as contributions from very large $p$ will be important at small $U$.

We can improve the numerics at small $U$ somewhat by employing a regularization procedure at small $U$.   Though we will not rigorously justify this procedure, we will check numerically that it gives results consistent with the more awkward (but manifestly correct) procedure described in the previous paragraph.  We begin by studying the leading terms in \eqref{eq:fC1} near $U=0$.  For $U>0$, Laurent expansion around $U=0$ gives
\begin{equation}
\label{eq:Expansion}
f\left(1,U;\Delta \right)=\frac{1}{32\pi {{U}^{3}}}+\frac{3-8\Delta +4{{\Delta }^{2}}}{64\pi U}+\frac{-2\Delta +3{{\Delta }^{2}}-{{\Delta }^{3}}}{6\pi }+O\left( U \right).
\end{equation}
We know that the singular terms should cancel when summed over $p \in {\mathbb Z}$ with a factor of $(-1)^p$.  This is especially natural for the first term on the right-hand side of \eqref{eq:Expansion} which is independent of $p$. Choosing to perform this sum using Dirichlet eta function regularization does indeed give zero as $\sum\limits_{p=1}^{\infty }{{{\left( -1 \right)}^{p}}} = - \eta (0)= -\frac{1}{2}$.

Using \eqref{eq:Delta}, the second term on the right-hand side of \eqref{eq:Expansion} becomes
\begin{equation}
\frac{3-8\Delta +4{{\Delta }^{2}}}{64\pi U}=\frac{3+4m^2\ell ^2+4\left(\frac{\ell }{R_{S^1}}\right)^2p^2}{64\pi U}.
\end{equation}
Thus, it gives a term independent of $p$ and a term proportional to $p^2$. Again applying Dirichlet eta-function regularization and recalling that $\sum\limits_{p=1}^{\infty }{{{\left( -1 \right)}^{p}p^2}}= - \eta(-2) = 0$, the $1/U$ term also cancels completely when summed over $p$.

Since we did not rigorously justify the use of eta-function regularization, there remains the possibility that we have missed some important finite piece that could remain after the above divergences cancel. But we now provide numerical evidence that this does not occur by computing $\hat f(1, 0;\ell/R_{S^1})$ in two different ways.  The first is to use \eqref{eq:Expansion} with Dirichlet regularization of the $1/U^3$ and $1/U$ terms and using Abel summation (i.e., replacing $(-1)^p$ by $(-1+\epsilon)^p$ and taking $\epsilon \rightarrow 0$) for the finite term.  The second is to compute the result for fixed but small $U\neq0$  by numerically summing over $p$ up to $|p|=N$ for some large $N$, but taking care to include an even number of terms with opposite signs; i.e., for $|U|>\alpha$ we take
\begin{equation}
\label{eq:summation}
\hat f_{\rm hybrid}\left( 1,U;\ell/R_{S^1}\right) =
f\left( 1,U;\Delta \left( 0 \right) \right)+2\sum\limits_{p=1}^{N-1}{{{\left( -1 \right)}^{p}}f\left( 1,U;\Delta \left( p \right) \right)}+{{\left( -1 \right)}^{N}}f\left( 1,U;\Delta \left( N \right) \right)\ \ \
\end{equation}
for some fixed large $N$.

\begin{figure}[t]
        \begin{center}
                \includegraphics[width=0.4\textwidth]{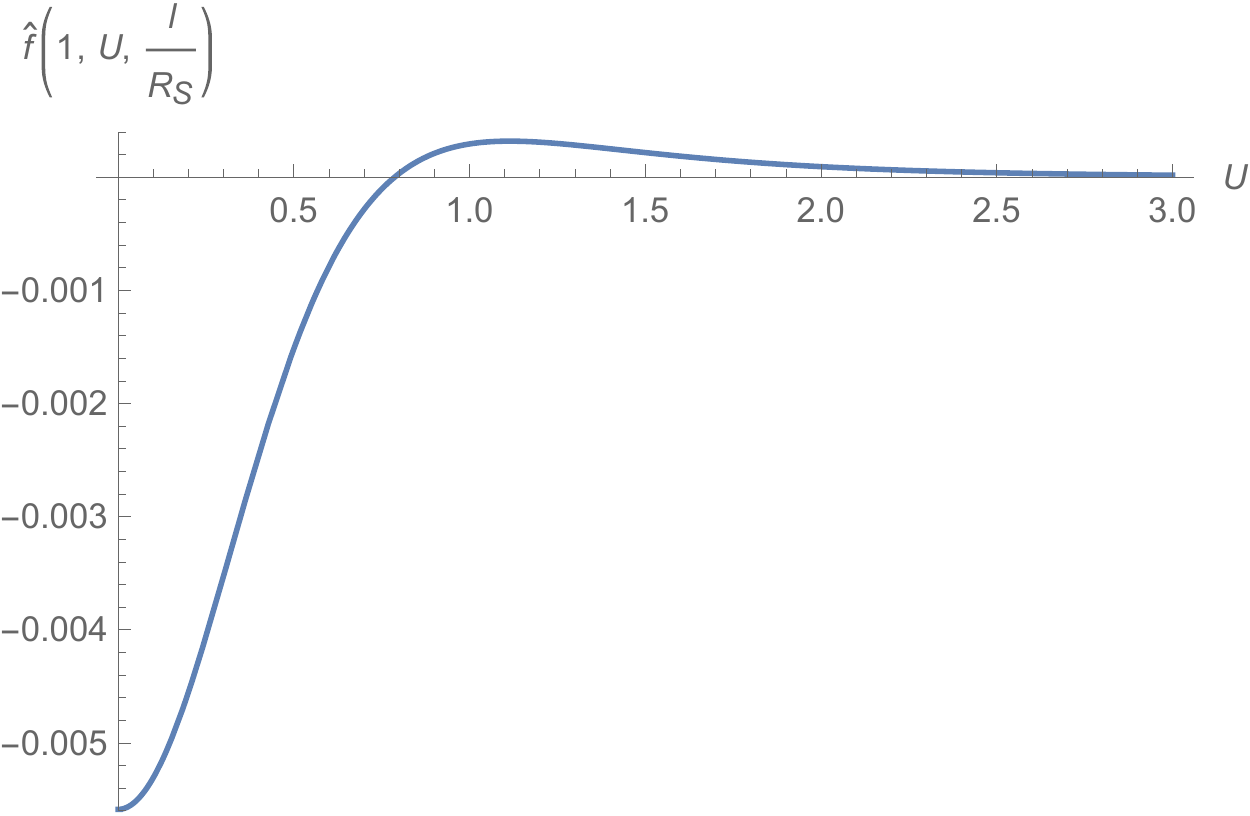}
                \includegraphics[width=0.4\textwidth]{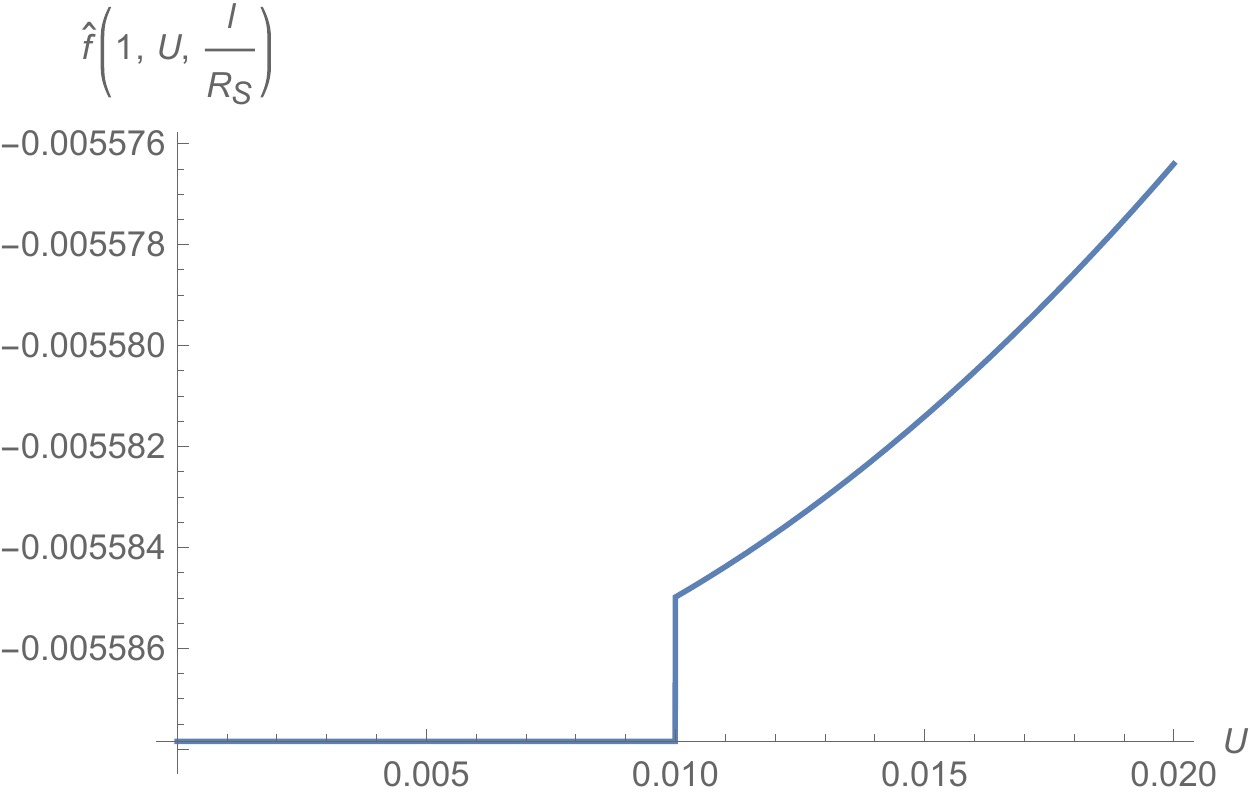}
        \end{center}
        \caption{An example of $\hat f_{\rm hybrid}\left(1,U; \ell/R_{S^1}\right)$ calculated numerically for $\ell/R_{S_1} =1$ and $\Delta \left( p \right)=1+\sqrt{1+{{p}^{2}}}$, $N=5000$, $\epsilon =10^{-6}$, and $\alpha =0.01$. The numerical integral gives $\int_{0}^{\infty }{\hat f\left(1,U;\ell /R_{S^1}\right)\text{d}U}=-9.05\times 10^{-4}$, which is negative.  In comparison, the triangle defined in the rightmost plot  by the horizontal line for $U< \alpha$ and the vertical line representing the discontinuity gives a measure of the numerical error in our computation of this integral and is of order $10^{-8}$.}
        \label{fig:2}
\end{figure}

Sample results are shown in figure \ref{fig:2}, where we plot $f_{\rm  hybrid}$ defined by introducing a parameter $\alpha >0$, performing the sums numerically for $|U|> \alpha$, and then taking $f_{\rm hybrid}$ to be constant for $|U|<\alpha$ with a value given by the above Abel summation. The small discontinuity at $U=\alpha$ in the resulting $\hat f_{\rm hybrid}$ supports the validity of the above regularization.  We can then approximate $\int \text{d}U \hat f(1,U; \ell/R_{S^1})$ by numerically integrating $\hat f_{\rm hybrid}$.

\begin{figure}[t]
        \begin{center}
                \includegraphics[width=0.4\textwidth]{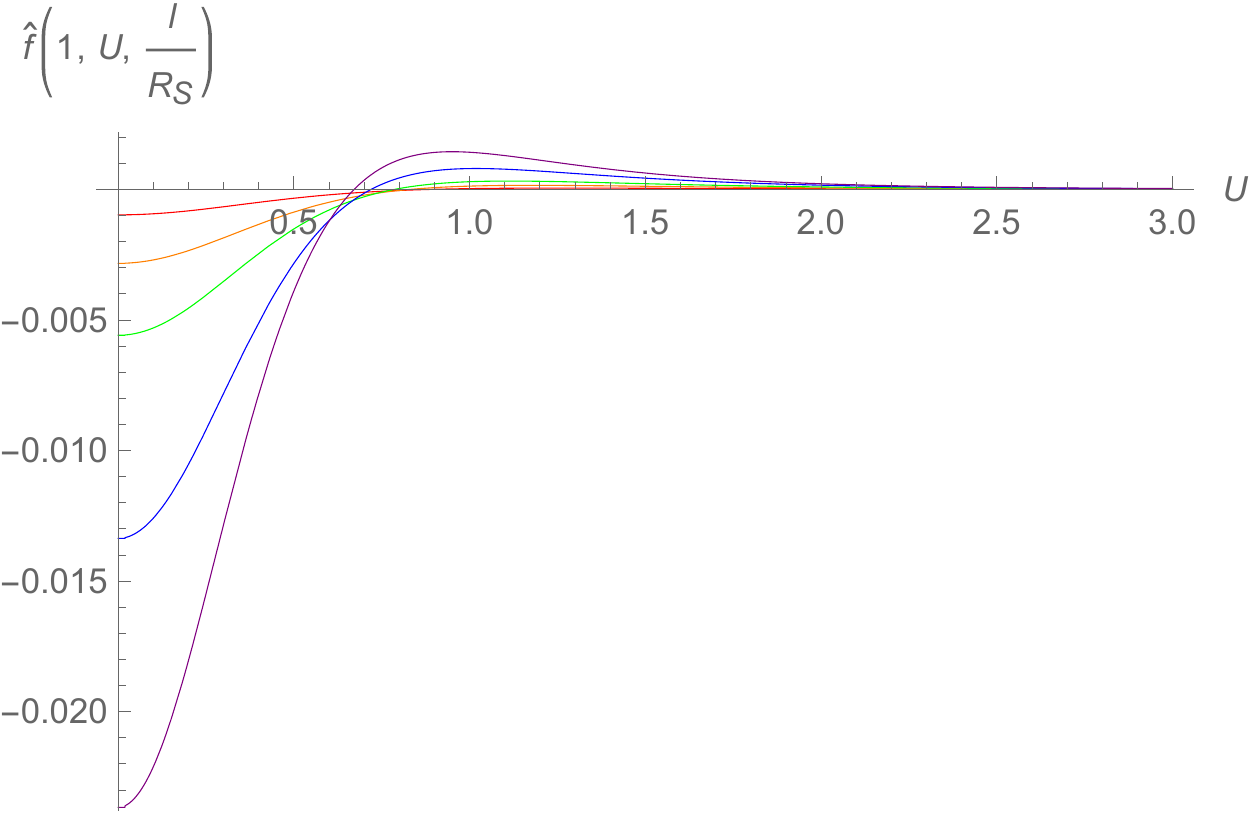}
                \includegraphics[width=0.4\textwidth]{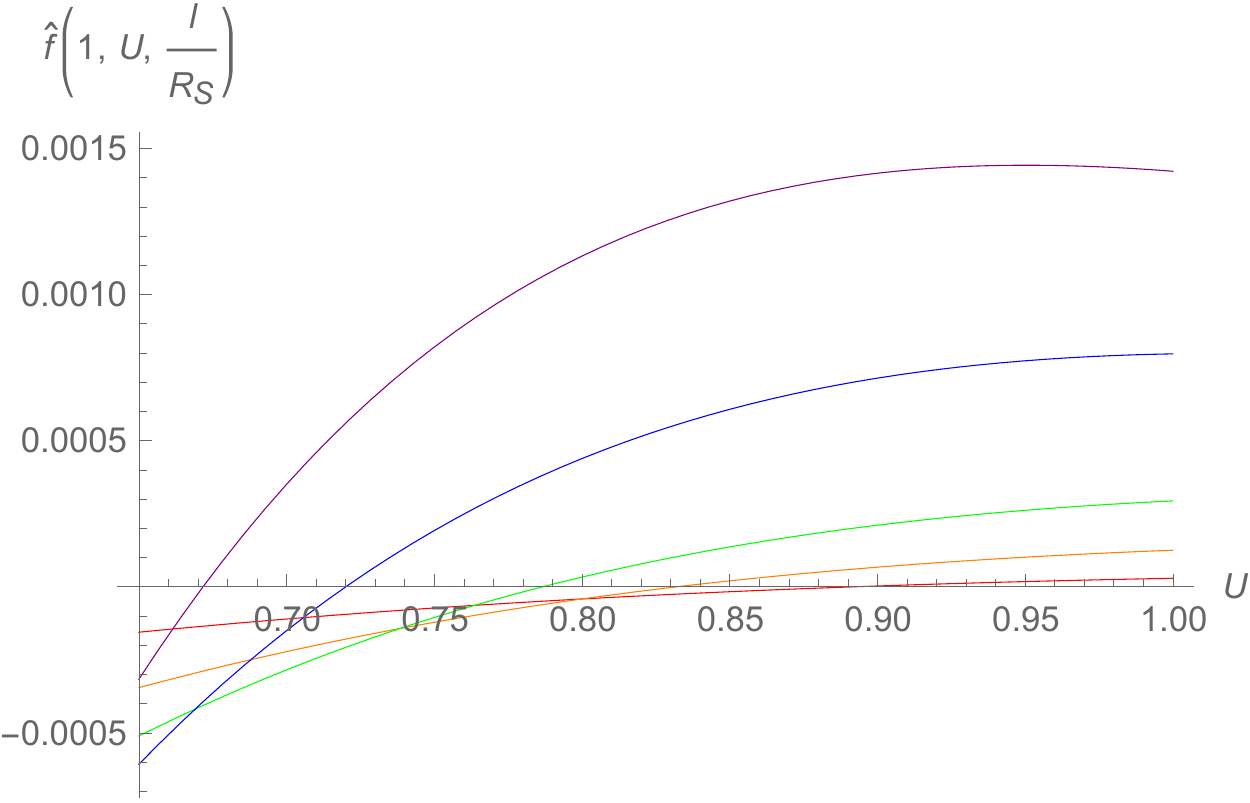}
                \end{center}
        \caption{Profiles of $\hat f$ for fixed $m=0$ and $\left(\ell/R_{S^1}\right)^2=0.5$ (red), $\left(\ell/R_{S^1}\right)^2=0.75$ (orange), $\left(\ell/R_{S^1}\right)^2=1$ (green), $\left(\ell/R_{S^1}\right)^2=1.5$ (blue), $\left(\ell/R_{S^1}\right)^2=2$ (purple). For this figure, we have chosen all ($\pm$) signs in \eqref{eq:Delta} to be ($+$) for all $p$. Note that the zero of $\hat f$ shifts to smaller $U$ as $\ell/R_{S^1}$ increases.}
        \label{fig:hatfb}
\end{figure}

\begin{figure}[t]
        \begin{center}
                \includegraphics[width=0.4\textwidth]{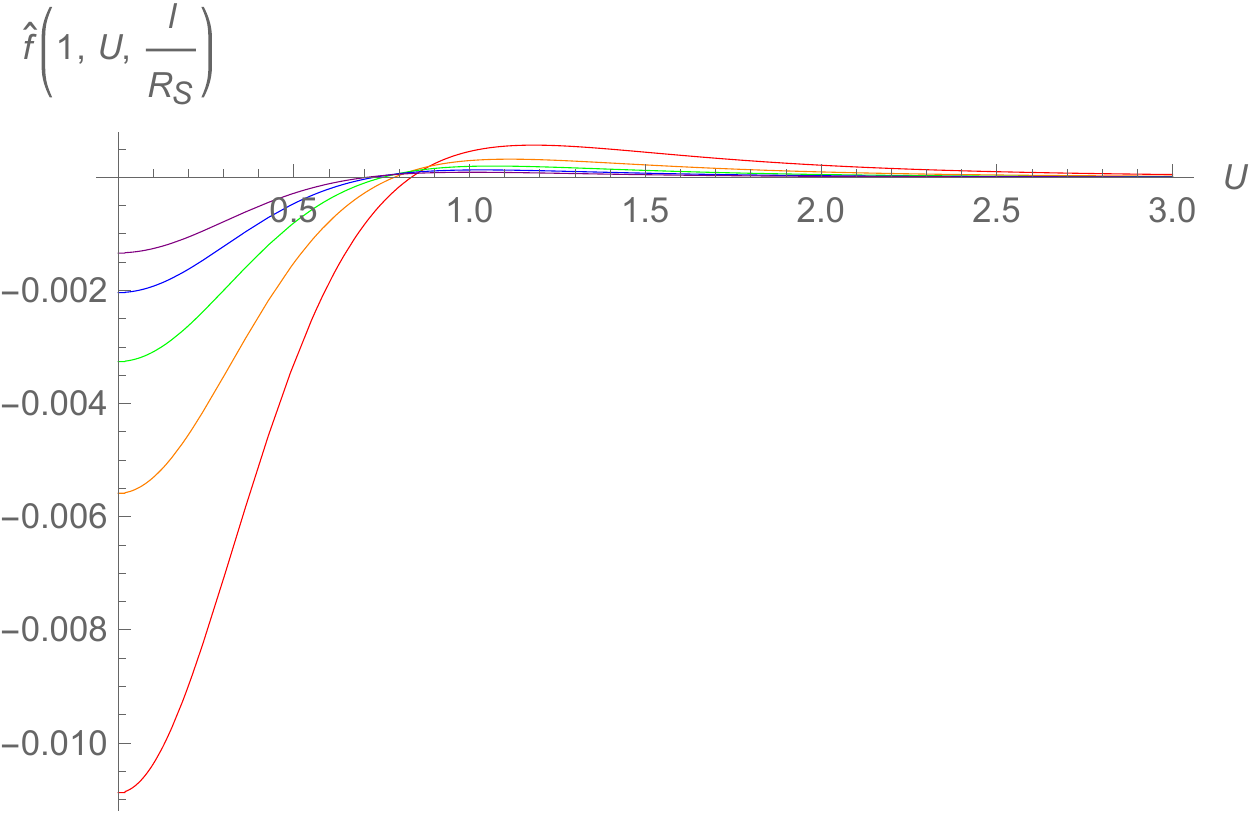}
                \includegraphics[width=0.4\textwidth]{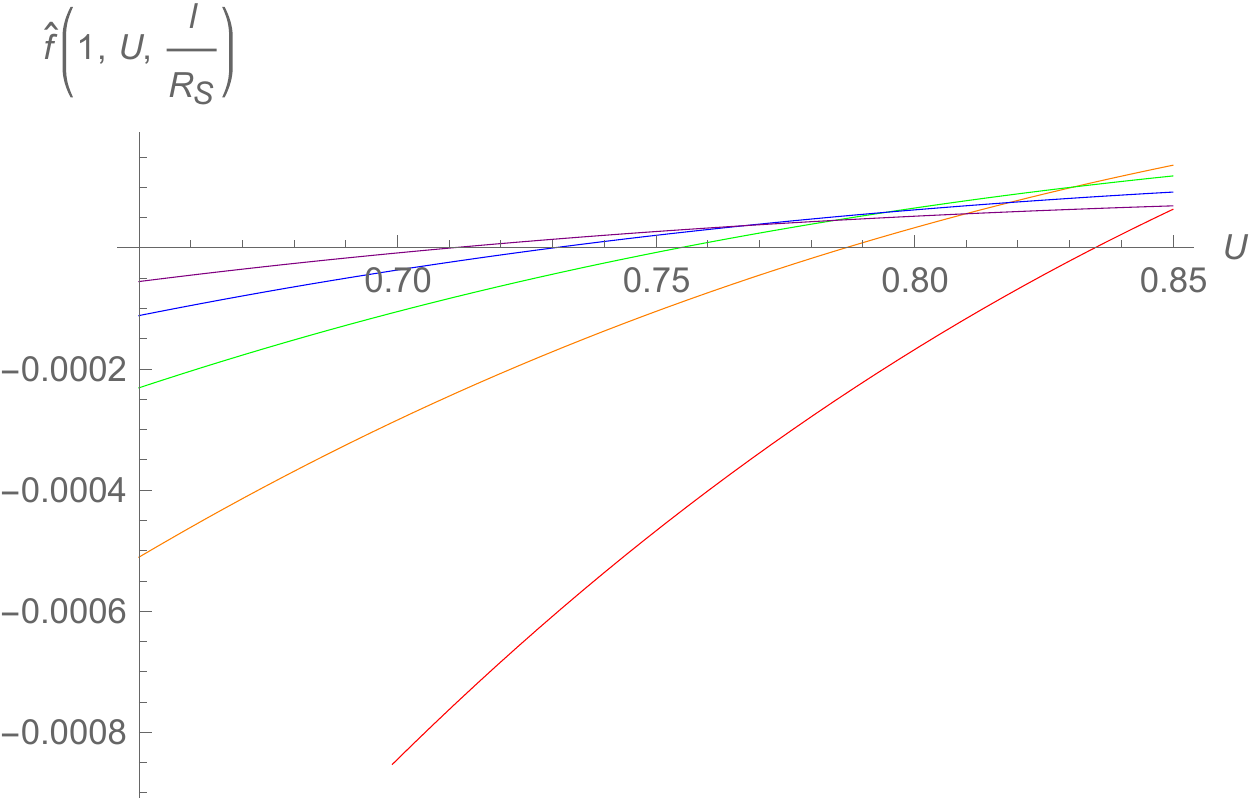}
                \end{center}
        \caption{Profiles of $\hat f$ for fixed $\ell/R_{S^1}=1$ and $m^2\ell ^2=-0.5$ (red), $m^2\ell ^2=0$ (orange), $m^2\ell ^2=0.5$ (green), $m^2\ell ^2=1$ (blue), $m^2\ell ^2=1.5$ (purple). For this figure, we have chosen all ($\pm$) signs in \eqref{eq:Delta} to be ($+$) for all $p$. As $m$ increases, the zero of $\hat f$ shifts to smaller $U$.}
        \label{fig:hatfa}
\end{figure}

It is interesting to compare the $\hat f$ in figure \ref{fig:2} with a graph of the first term $f(1,U,\Delta =2)$ in its definition \eqref{eq:pSum} (the orange curve on the right figure of figure \ref{fig:f}). The first term is manifestly positive, while the intgeral of $\hat f$ is negative.  This emphasizes the importance of the higher terms in the sum near $U=0$.  The dependence of $\hat f$ on $\Delta(p=0)$ and $\ell/R_{S^1}$ is illustrated in figures \ref{fig:hatfb} and \ref{fig:hatfa}.

\begin{figure}[t]
        \begin{center}
                \includegraphics[width=0.4\textwidth]{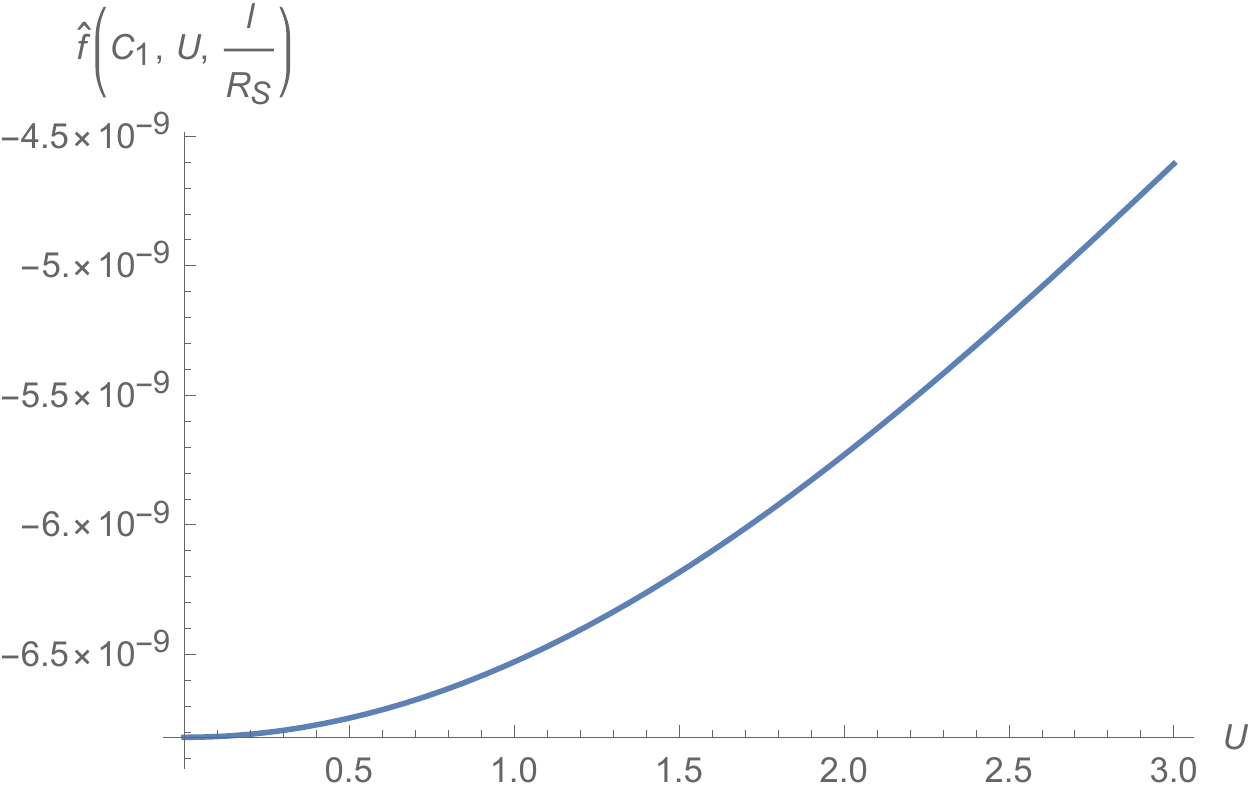}
                \includegraphics[width=0.4\textwidth]{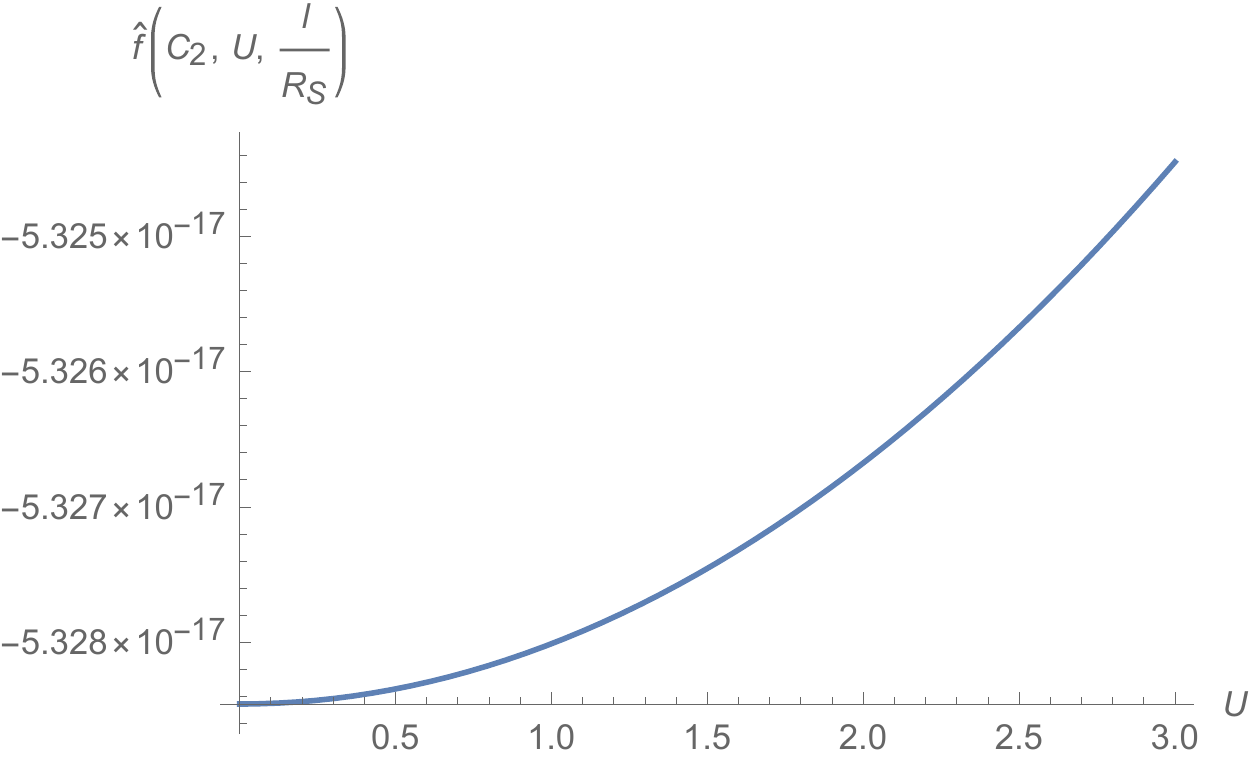}
                \end{center}
        \caption{The $n=1,2$ contributions for the example in figure \ref{fig:2}, where we have chosen $r_+/\ell =1$. We find $\int_{0}^{\infty }{\hat f\left(C_1,U;\ell /R_{S^1}\right)\text{d}U}=-1.18\times 10^{-8}$ (left) and $\int_{0}^{\infty }{\hat f\left(C_2,U;\ell /R_{S^1}\right)\text{d}U}=-2.10\times 10^{-15}$ (right). Thus, for black holes with size $r_+/\ell \ge 1$, contributions from $n\neq 0$ terms are negligible.}
        \label{fig:highern}
\end{figure}

\begin{figure}[t]
        \begin{center}
                \includegraphics[width=0.4\textwidth]{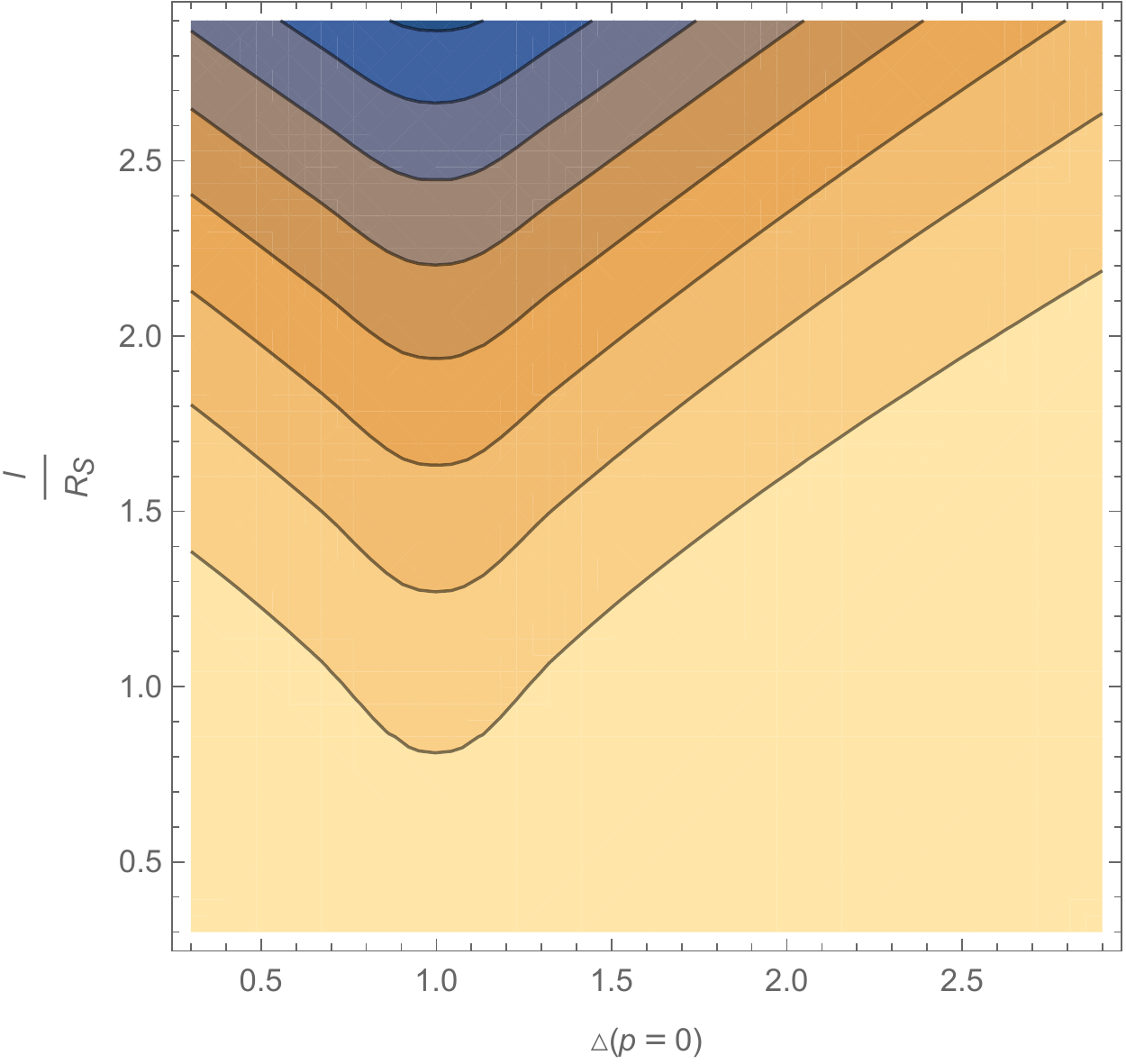}
                \includegraphics[width=0.08\textwidth]{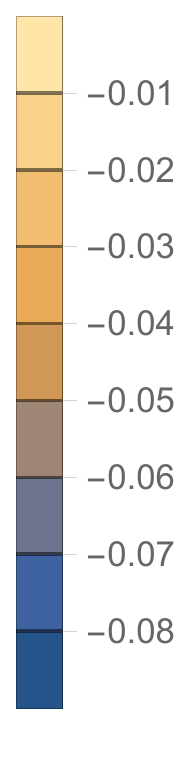}
                \end{center}
        \caption{The quantity $\int _0^\infty \text{d}U \hat f(1,U; \ell/R_{S^1})$ as a function of $\Delta(p=0)$ and $\ell/R_{S^1}$. For all $p\neq 0$, we have chosen all ($\pm$) signs in \eqref{eq:Delta} to be ($+$).}
        \label{fig:3}
\end{figure}

In the limit of large $r_+/\ell $, the contributions from $n\neq 0$ are suppressed and the $\left<T_{kk}\right>$ exactly becomes $\hat f\left(1,U;\ell /R_{S^1}\right)$. Moreover, numerical calculation shows that this is a good approximation even for $r_+/\ell \ge 1$; see figure \ref{fig:highern}. Up to the factor of $4\pi G$ in \eqref{eq:DeltaVfinal}, $\Delta V$ becomes just $\int \text{d}U \hat f(1,U; \ell/R_{S^1})$.  Numerical results for this integral are shown in figure \ref{fig:3} with the signs $(\pm)$ in \eqref{eq:Delta} chosen to be $(+)$ for $p\neq 0$.  The integral is negative for all such cases we have explored.  As one would expect, the magnitude of the integral becomes large for large $\ell/R_{S^1}$.  We again find a finite (negative) shift $\Delta V$ at $r_+ = \infty$ for $\Delta(p=0)=0$, and $\Delta V$ vanishes in the limit of large mass $m$, though the maximum value of $|\Delta V|$ depends on $\ell/R_{S^1}$.

\begin{figure}[t]
        \begin{center}
                \includegraphics[width=0.4\textwidth]{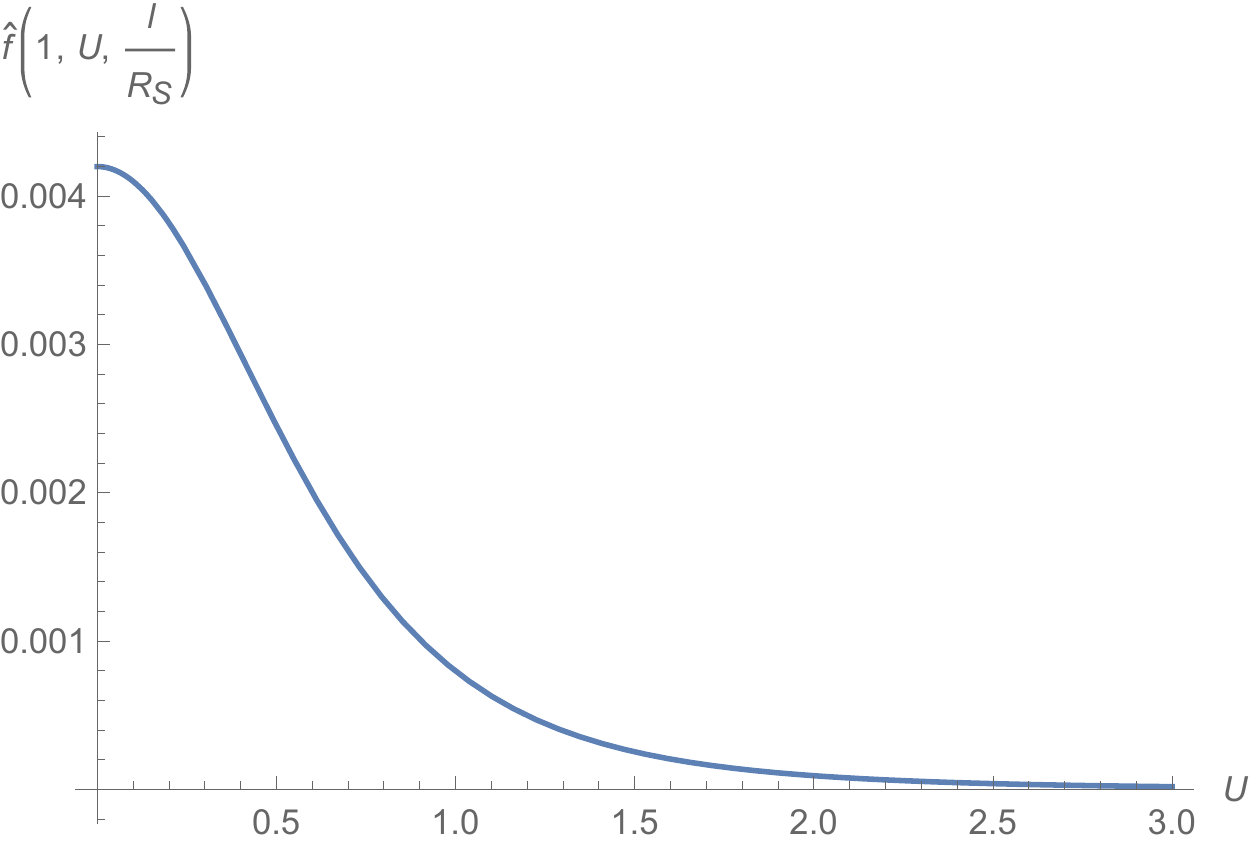}
                \includegraphics[width=0.4\textwidth]{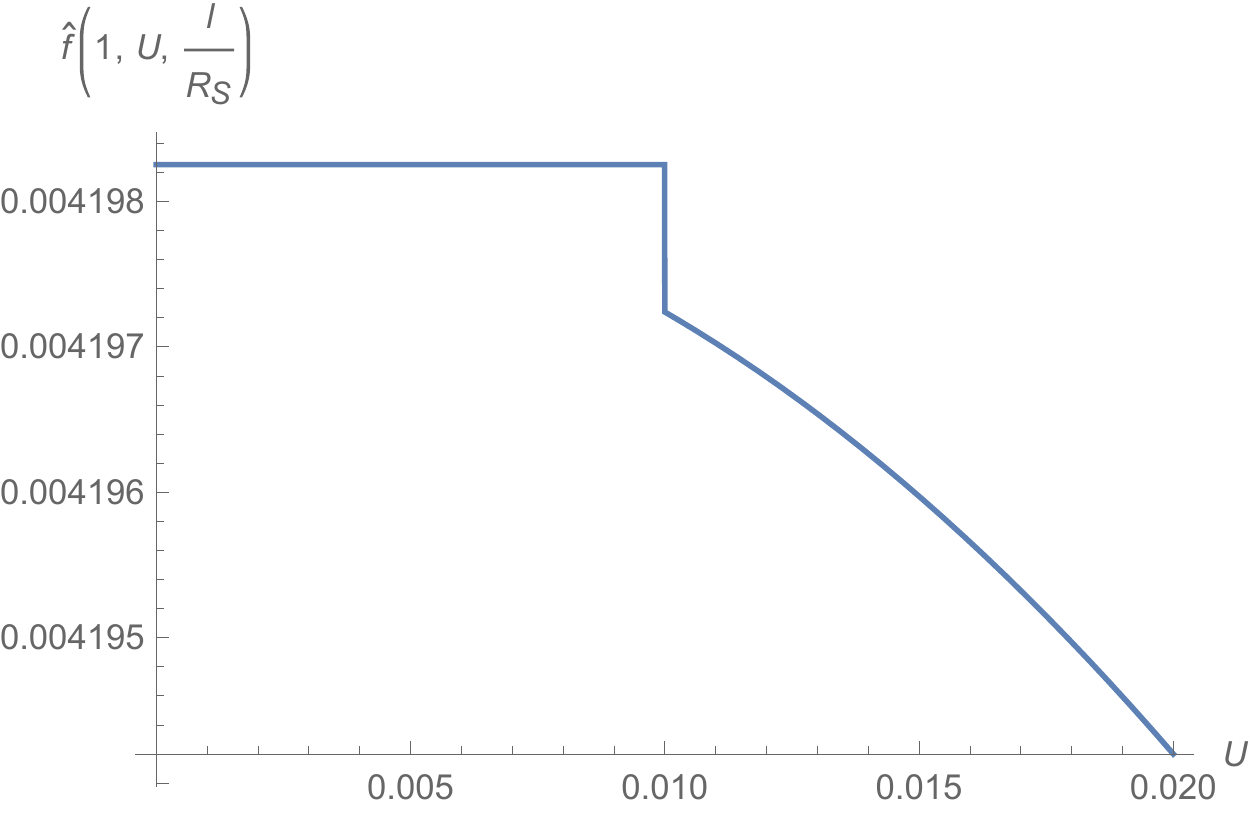}
        \end{center}
        \caption{An example of $\hat f_{\rm hybrid}\left(1,U;\ell/R_{S^1}\right)$ that is everywhere positive. Here $m^2\ell ^2=-0.12$, $\left(\ell/R_{S_1}\right)^2=0.1$, $N=5000$, $\epsilon =10^{-6}$, and $\alpha =0.01$. Moreover, we have chosen the ($\pm$) signs in \eqref{eq:Delta} to be ($-$) for $|p|=1$ and to be ($+$) for both $p=0$ and $|p| > 1$. We find $\int_{0}^{\infty }{\hat f\left(1,U;\ell /R_{S^1}\right)\text{d}U}=1.44\times 10^{-3} > 0$.}
        \label{fig:nontraversable}
\end{figure}

However, it turns out that for some choices $m$ and $\ell /R_{S^1}$, we can choose the $(\pm )$ signs in \eqref{eq:Delta} to be $(-)$ for $|p|=1$ and to be $(+)$ for all other values of $p$ (including $p=0$).  In at least some such cases $\int_{0}^{\infty }{\hat f\left(1,U;\ell /R_{S^1}\right)\text{d}U}$ is positive and the back-reacted wormhole remains non-traversable when our scalar satisfies periodic boundary conditions. One example is shown in figure \ref{fig:nontraversable}.

\section{Rotating traversable wormholes with Kaluza-Klein zero-brane orbifolds}
\label{sec:KKZBO}

We now turn to a slightly more complicated construction that allows rotation and thus admits a smooth extremal limit.  We begin with the rotating BTZ metric
\begin{equation}
\label{eq:rotBTZ}
  \text{d}s^2 = \frac{1}{\left(1+ UV\right)^2} \left\{-4\ell^2 \text{d}U \text{d}V +4\ell r_- \left(U \text{d}V- V\text{d}U \right)\text{d}\phi +\left[r_+^2 \left(1-U V\right)^2+4U V r_-^2 \right] \text{d}\phi^2\right\}.
\end{equation}
Note that the ${\mathbb Z}_2$ operations used earlier exchange $U \leftrightarrow V$ while preserving the sign of $\text{d}\phi$.  As a result, they change the sign of the $4 r_- \left(U \text{d}V -V \text{d}U\right)\text{d}\phi$ term in \eqref{eq:rotBTZ} and are not isometries for $r_- > 0$.

This can be remedied by simultaneously acting with $\phi \rightarrow - \phi$.  To remove the would-be fixed-points at $\phi = 0, \pi$ for $U=V=0$, as for the KKEOW brane, we consider a Kaluza-Klein setting involving BTZ $\times S^1$ and act on this circle with the antipodal map $\theta \rightarrow \theta+\pi$.  Our full isometry is thus $J: (U,V, \phi, \theta) \rightarrow (V,U,-\phi, \theta +\pi)$.  This quotient breaks rotational symmetry by singling out the points $\phi =0, \pi$ as Klauza-Klein orbifolds (i.e., as points that become oribifold singularities  with deficit angle $\pi$ after Kaluza-Klein reduction along the internal $S^1$), but allows non-zero rotation and admits a smooth extremal limit.  The computations then proceed much as before, though we review the main points below.

\subsection{Geometry and back-reaction}

At first order in the metric perturbation $h_{ab}$, the analysis of null geodesics traversing the wormhole turns out to be identical to that in the non-rotating case; i.e., equations \eqref{eq:geodesic} and \eqref{eq:gengeo} continue to hold without change.  However, choosing a conformal frame in which the boundary metric is $\text{d}s^2_{\partial BTZ} = - \text{d}t^2 + \ell^2 \text{d}\phi^2$ now yields
\begin{equation}
\label{eq:trotV}
t = \pm \frac{\ell^2r_+}{2(r_+^2-r_-^2)}\ln \left(\pm \frac{V}{\ell}\right),
\end{equation}
with the signs being both $(+)$ on the right boundary and both $(-)$ on the left.

Nevertheless, the critical change occurs in
the linearized Einstein equation that determines $h_{ab}$.   We find
\begin{equation}
\begin{aligned}
8 \pi G \langle T_{kk}\rangle= & -\frac{1}{2 \ell^2 r_+^2}\left[  \left(r_-^2-r_+^2\right) h_{kk}+2 \ell r_- \partial_\phi h_{kk}+\ell^2\partial_\phi^2h_{kk} \right. \\
& \left. +\left(r_-^2-r_+^2\right) \partial_U\left(U h_{kk}\right)-2 \ell^2 \partial_U\partial_\phi h_{k\phi}+\ell^2\partial_U^2h_{\phi \phi }\right],
\end{aligned}
\end{equation}
where $G$ is the 3 dimensional Newton's constant. Integrating over $U$ and applying asymptotically AdS boundary conditions gives
\begin{equation}
\label{eq:introt}
 8 \pi G \int \langle T_{kk}\rangle \text{d}U=-\frac{1}{2\ell^2 r_+^2}\left[ \left(r_-^2 - r_+^2\right) + 2\ell r_- \partial_\phi + \ell^2 \partial^2_\phi \right]\int h_{kk} \text{d}U.
\end{equation}
Equation \eqref{eq:introt} is easily solved for $\left(\int \text{d}U h_{kk} \right)(\phi)$ using a Green's function $H$, so that
\begin{equation}
\label{eq:rothkk}
\left(\int \text{d}U h_{kk}\right)\left(\phi \right) =8 \pi G  \int \text{d} \phi' H\left (\phi - \phi '\right ) \int \text{d}U \langle T_{kk}\rangle \left(\phi ' \right), \end{equation}
with
\begin{equation}
\label{eq:Hphi}
H(\phi - \phi') =
\begin{cases}
      \dfrac{e^{-(r_+-r_-) (\phi'-\phi)/\ell}}{2 r_+ /\ell\left[1-e^{-2\pi  (r_+-r_-)/\ell}\right]}+\dfrac{e^{(r_-+r_+) (\phi'-\phi )/\ell}}{2 r_+ /\ell\left[e^{2\pi  (r_-+r_+)/\ell}-1\right]} & \phi' \geq \phi \\
     \left.\dfrac{e^{(r_-+r_+) (2\pi -\phi+\phi')/\ell}}{2 r_+ /\ell\left[e^{2\pi  (r_-+r_+)/\ell}-1\right]}+\dfrac{e^{-(r_+-r_-) (2\pi -\phi+\phi')/\ell}}{2 r_+ /\ell\left[1-e^{-2\pi (r_+-r_-)/\ell}\right]} \right. & \phi' \leq \phi .\\
   \end{cases}
\end{equation}
in position space where we take $\phi, \phi' \in [0,2\pi)$. It is also useful to write $H$ in Fourier space:
\begin{equation}
\label{eq:Hphi2}
H\left(\phi - \phi ' \right) =\sum_q e^{iq\left(\phi-\phi'\right)} H_q, \ \ \ H_q = \frac{1}{2\pi} \frac{2\ell^2 r_+^2}{r_+^2-r_-^2 -2iq\ell r_- + \ell^2 q^2}.
\end{equation}
Note in particular that the zero-mode Green's function $H_{q=0} = \frac{\ell^2 r_+^2}{\pi(r_+^2-r_-^2)}$ diverges in the extremal limit $r_- \rightarrow r_+$. This feature was also independently and simultaneously noted in \cite{Caceres:2018ehr}, where the somewhat different form of their expression appears to be due to differences in the detailed definition of the Kruskal-like coordinates.   While we have not explored the connection in detail, it is natural to expect this feature to be related to other known instabilities of extreme black holes \cite{Marolf:2010nd,Marolf:2011dj,Aretakis:2011ha,Aretakis:2011hc,Aretakis:2011gz,Aretakis:2012ei,Lucietti:2012sf}
and in particular to the Aretakis instability for gravitational perturbations (see e.g. \cite{Lucietti:2012sf} for the Kerr case), though our present instability seems to occur only for the zero-mode while at least in Kerr the Aretkis instability is strongest at large angular momentum \cite{Casals:2016mel}.    In our first-order perturbative analysis, this divergence implies that any non-vanishing zero-mode component $\int \text{d}U \text{d}\phi \langle T_{kk}\rangle$ of the averaged null stress tensor  in the extremal limit leads to diverging $\Delta V$.  The perturbative analysis can then no longer be trusted in detail, though for $\int \text{d}U \text{d}\phi \langle T_{kk}\rangle <0$ it certainly suggests that the wormhole remains traversable until very late times $V_f$.  And so long as
$V_f > \ell,$ the extreme limit of \eqref{eq:trotV} then implies that the wormhole remains traversable at arbitrarily late times $t$; i.e., it becomes an eternal static wormhole.

In contrast, the non-zero modes of $H_q$ remain finite at extremality.  So even though the source
$\int \text{d}U \langle T_{kk}\rangle$ will break rotational symmetry, in the extreme limit the geometry approximately retains this invariance and it suffices to study only the zero mode.
Recalling that the BTZ temperature is given by $T = \frac{r_+^2 - r_-^2}{2 \pi r_+ \ell^2 }$, we may write
\begin{equation}
\frac{T \pi}{r_+}\int h_{kk} \text{d}U\text{d}\phi  = 8 \pi G \int \langle T_{kk}\rangle \text{d}U \text{d}\phi,
\end{equation}
so that \eqref{eq:geodesic} gives
\begin{equation}
\label{eq:TDVAv}
 T \Delta V_{\rm average} = \frac{2 G r_+}{\pi \ell^2} \int_{-\infty}^\infty \int_{-\pi/2}^{\pi/2} \langle T_{kk}\rangle \text{d}U \text{d}\phi.
\end{equation}

This is a convenient form for displaying results in the extreme limit, which will be the main focus of our calculations below.  And more generally if $\Delta V_{\rm average} < 0$ it follows that the wormhole must become traversable when entered from at least one direction.  However,  it is also interesting to consider the high temperature limit $r_+ \rightarrow \infty$ (say, for $r_-=0$) in which the Green's function $H(\phi-\phi')$ becomes sharply peaked at $\phi-\phi'=0$ and the $\int \text{d}U h_{kk}$ at each $\phi$ can be thought of as locally determined by $\int \text{d}U \langle T_{kk}\rangle$.

\subsection{KKZBO results}

We again compute the stress tensor using \eqref{eq:crossterm} and the BTZ Green's function \eqref{eq:GBTZ}, which remains valid so long as we use the correct expression for proper distance in the rotating BTZ metric
\begin{equation}
\begin{aligned}
\sigma(x,x_n{}')
=&\frac{1}{(U V+1) (U' V'+1)}2 \ell ^2 \left\{(U V-1) (U' V'-1) \cosh \left[r_+ (\phi -\phi_n')\right]\right. \\ &+2 \cosh \left[r_- (\phi -\phi_n')\right] (U V'+V U')-2 UV' \sinh \left[r_- (\phi -\phi_n')\right]\\
&\left. +2 V U' \sinh \left[r_- (\phi -\phi_n')\right]-U U' V V'-U V-U' V'-1\right\}
\end{aligned}
\end{equation}
As before, the basic elements of our computations are the functions
\begin{equation}
\label{eq:integrandrot}
f(C_+,C_-,S_-,U; \Delta):=
\langle 0_{HH,{\rm AdS}_3}|
\partial_U \phi(x) \partial_U \phi(x')
| 0_{HH,\rm{AdS}_3}\rangle|_{V=0}
\end{equation}
defined by the vacuum on global AdS$_3$ where the dependence on angles appears only through $C_\pm = \cosh(r_\pm[\phi-\phi'])$ and $S_- = \sinh(r_-[\phi-\phi'])$. We find
\begin{equation}
\begin{aligned}
  f\left( {{C}_{-}},{{C}_{+}},{{S}_{-}},U;\Delta  \right)=&\frac{{{\left( \sqrt{{{Y}^{2}}-1}+Y \right)}^{-\Delta }}\left(S_-+C_-\right)}{2\pi {{\left( {{Y}^{2}}-1 \right)}^{5/2}}}\Bigg[\left(1-Y^2\right)\bigg(1+\Big(Y^2-1 \\
  &+ Y \sqrt{Y^2-1}\Big)\Delta \bigg) +2(S_-+C_-)U^2 \bigg((2-\Delta ) \Delta  \sqrt{Y^2-1}\\
  &+\Delta  (\Delta +1) Y^3-\left(\Delta ^2+\Delta -3\right) Y+\Delta  (\Delta +1) \sqrt{Y^2-1} Y^2\bigg)\Bigg],
\end{aligned}
\end{equation}
where $Y\equiv 2{{U}^{2}}\left( {{C}_{-}}+{{S}_{-}} \right)+{{C}_{+}}$. Much as in section \ref{sec:EOTW} we write
\begin{equation}
\label{eq:nSumrot}
\left\langle {{T}_{kk}}\left( U \right) \right\rangle =\sum\limits_{n\in \mathbb{Z}}{\hat f\left( {{C}_{+n}},C_{-n},S_{-n},U;\ell/R_{S^1}\right)}
\end{equation}
where
\begin{equation}
\begin{aligned}
\label{eq:pSumrot}
\hat f\left({{C}_{+n}},C_{-n},S_{-n},U;\ell/R_{S^1}\right)=&f\left({{C}_{+n}},C_{-n},S_{-n},U;\Delta \left( 0 \right) \right)\\
&+2\sum\limits_{p=1}^{\infty }{{{\left( -1 \right)}^{p}}f\left( {{C}_{+n}},C_{-n},S_{-n},U;\Delta \left( p \right) \right)}
\end{aligned}
\end{equation}
with $\phi'_n = 2\pi n - \phi$.

At general values $\phi\neq 0,\pi$ we have $C\neq 1$ and each term above is separately finite and smooth.  The same is true at $\phi=0,\pi$ for $n \neq 0$.  But for $n=0$ and $\phi =0,\pi$, the contribution for each $p$ diverges at $U=0$.  In fact, since $C_\pm = \cosh[r_\pm(\phi-\phi')] =1$, $S_- = \sinh[r_-(\phi-\phi')] =0$, at $n=0$, $\phi=0,\pi$ we find $\hat f\left( {{C}_{+n}},C_{-n},S_{-n},U;\ell/R_{S^1}\right)= \hat f\left( 1,U;\ell/R_{S^1}\right)$ ; i.e., in this case the computations reduce precisely to those for the $n=0$ term studied for the non-rotating KKEOW brane in section \ref{sec:EOTW}.

\begin{figure}
\centering
\begin{subfigure}{.45\linewidth}
\centering
\includegraphics[width=1\linewidth]{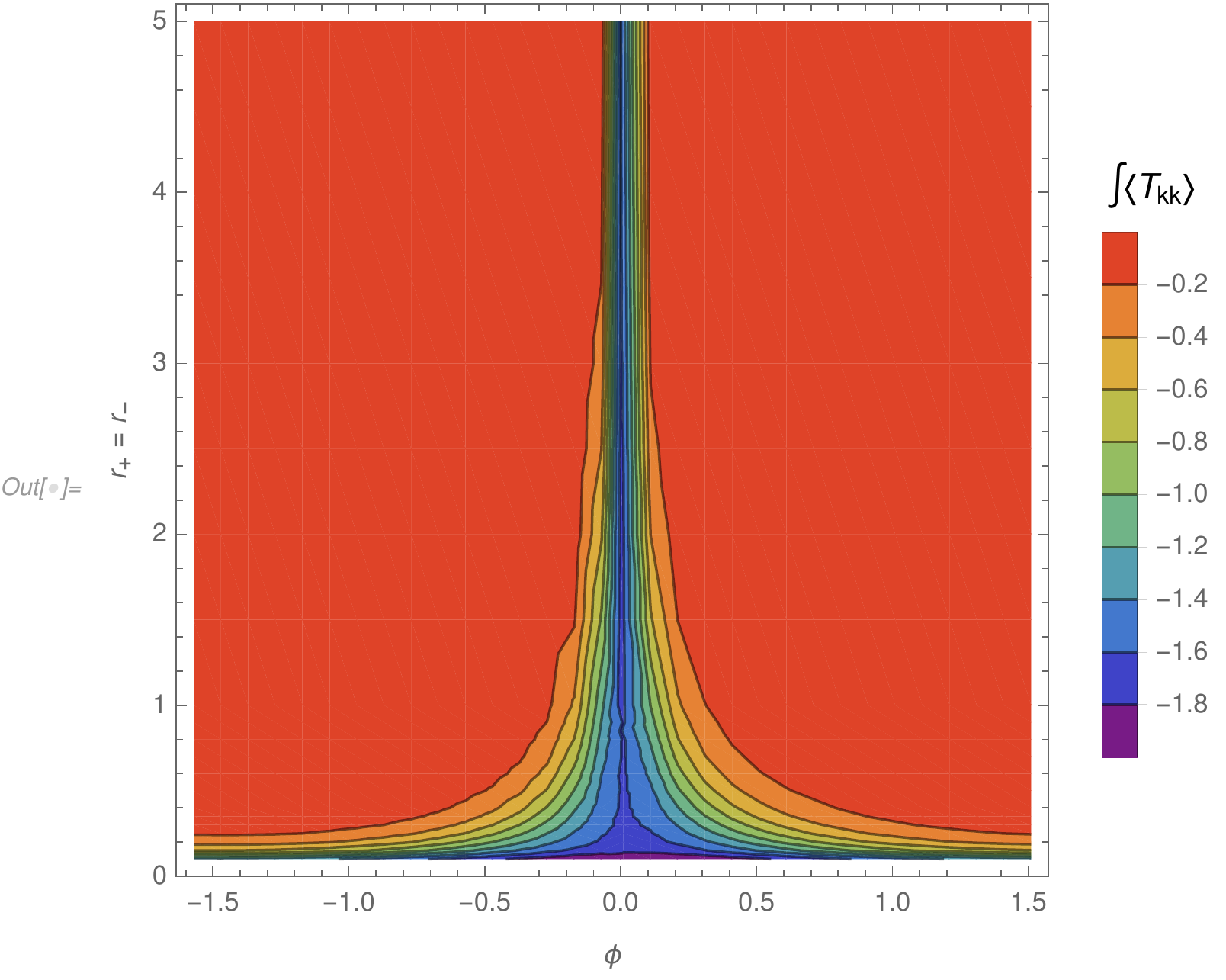}\quad%

\end{subfigure}
\begin{subfigure}{.46\linewidth}
\centering
\includegraphics[width=1\linewidth]{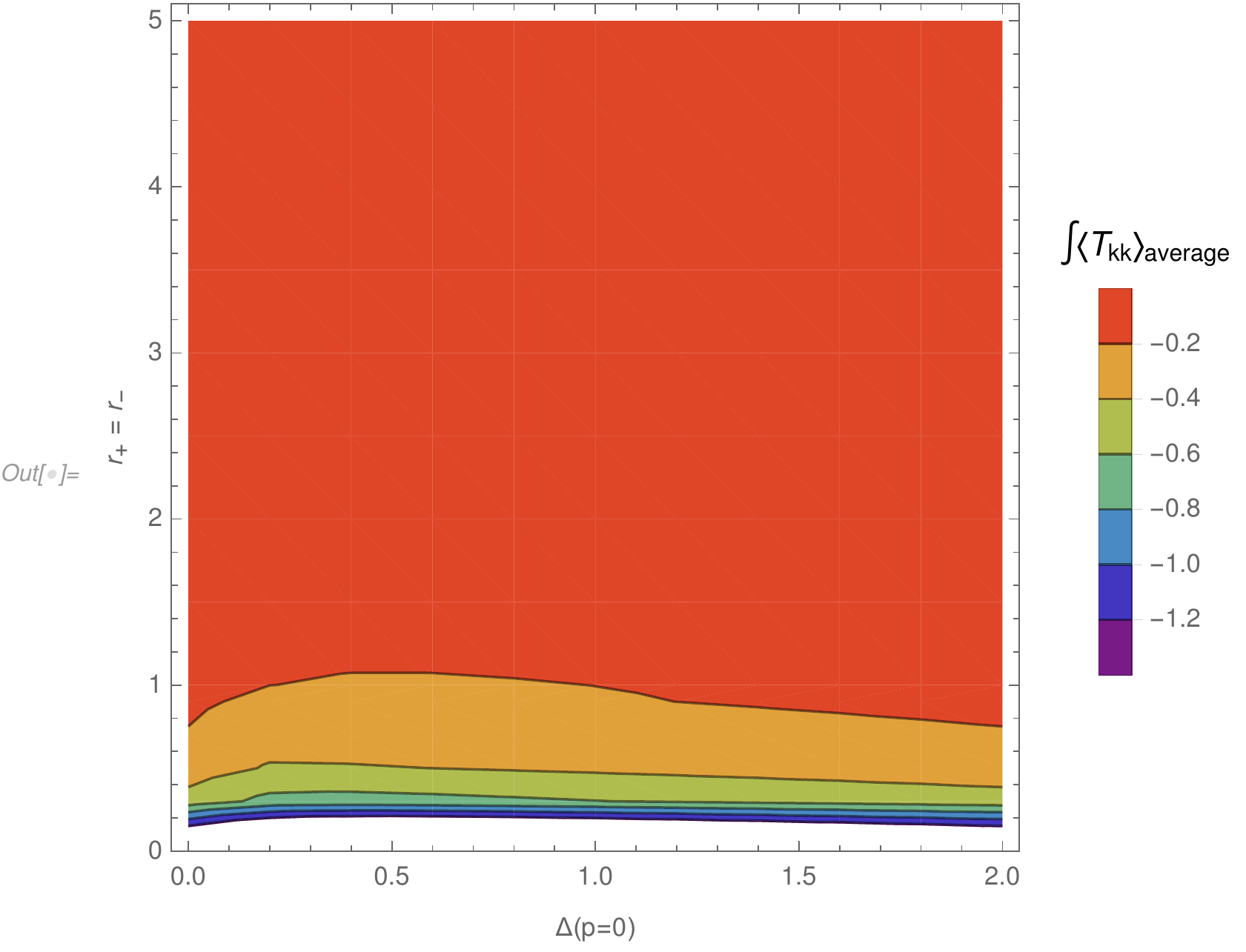}\quad%

\end{subfigure}
\caption{{\bf Left:} The dependence of $\int \langle T_{kk+}\rangle \text{d}U$ on $\phi$ and $r_+$ at extremality ($r_- = r_+$, in units of $\ell$). {\bf Right:} Dependence of the zero-mode $\int \langle T_{kk+} \rangle \text{d}Ur \text{d}\phi$ on $\Delta(p =0)$ (with all $\pm$ signs in \eqref{eq:Delta} chosen to be $+$ for $p\neq 0$) and $r_- = r_+$ in units of $\ell$. For both figures, we have chosen $\ell/R_{S^1} = 10$ and $m=0$.}
\label{fig:rotatingcontours}
\end{figure}

\begin{figure}
\centering
\includegraphics[width=0.5\linewidth]{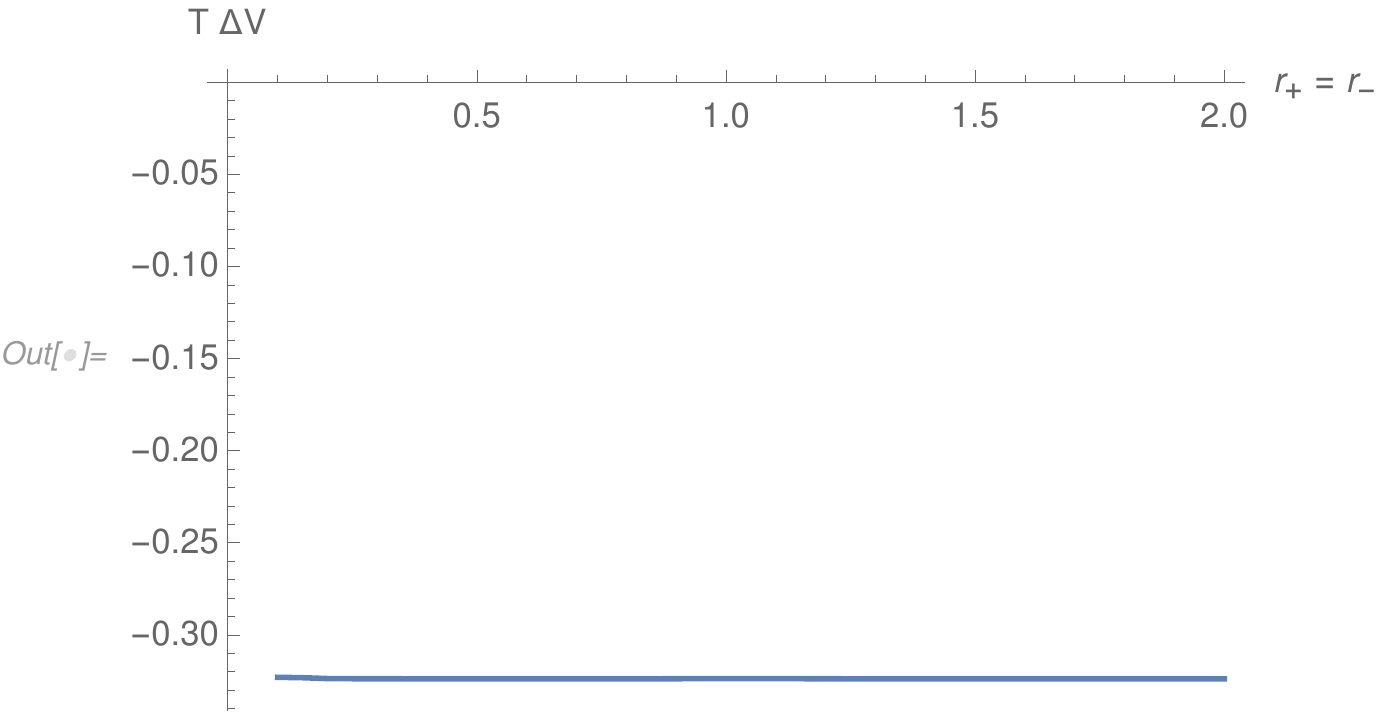}
\caption{$\Delta V_{\rm average} T$ at extremality as a function of $r_- = r_+$, with $\ell = G=1$. For this figure, we have chosen $\Delta(p=0) = 2$ and $\ell/R_{S^1} = 10$.  Though we have not peformed a thorough analysis of numerical errors, our results appear consistent with this quantity perhaps being independent of $r_+ = r_-$.}
\label{fig:DVT}
\end{figure}

Numerical results computed using a function $\hat f_{\rm hybrid}$ analogous to that in section \ref{sec:EOTW} are displayed in figures \ref{fig:rotatingcontours} and \ref{fig:DVT}.
As for the EOW brane, the analysis simplifies in the limit of large $r_+/\ell $ where contributions from $n \neq 0$ can be ignored.  In that limit, the stress tensor profile becomes sharply peaked near $\phi=0,\pi$ on a scale set by the Kaluza-Klein scale and the mass of the scalar field (though in a manner that is not symmetric under $\phi \rightarrow -\phi$); see figure \ref{fig:rotatingcontours} (left). As shown in figure \ref{fig:rotatingcontours} (right), the integral of the stress tensor becomes large (and negative) at small values of $R_{S^1}$, corresponding to the fact that Kaluza-Klein reduction on the $S^1$ gives orbifold singularities at which the stress tensor would diverge.  But the back-reaction \eqref{eq:TDVAv} involves an extra factor of $r_+$ and, as shown in figure \ref{fig:DVT}, our numerics for the quantity $\Delta V T$ suggest that this quantity may become independent of $r_+=r_-$ in the extremal limit.

\begin{figure}
\centering
\includegraphics[width=0.4\linewidth]{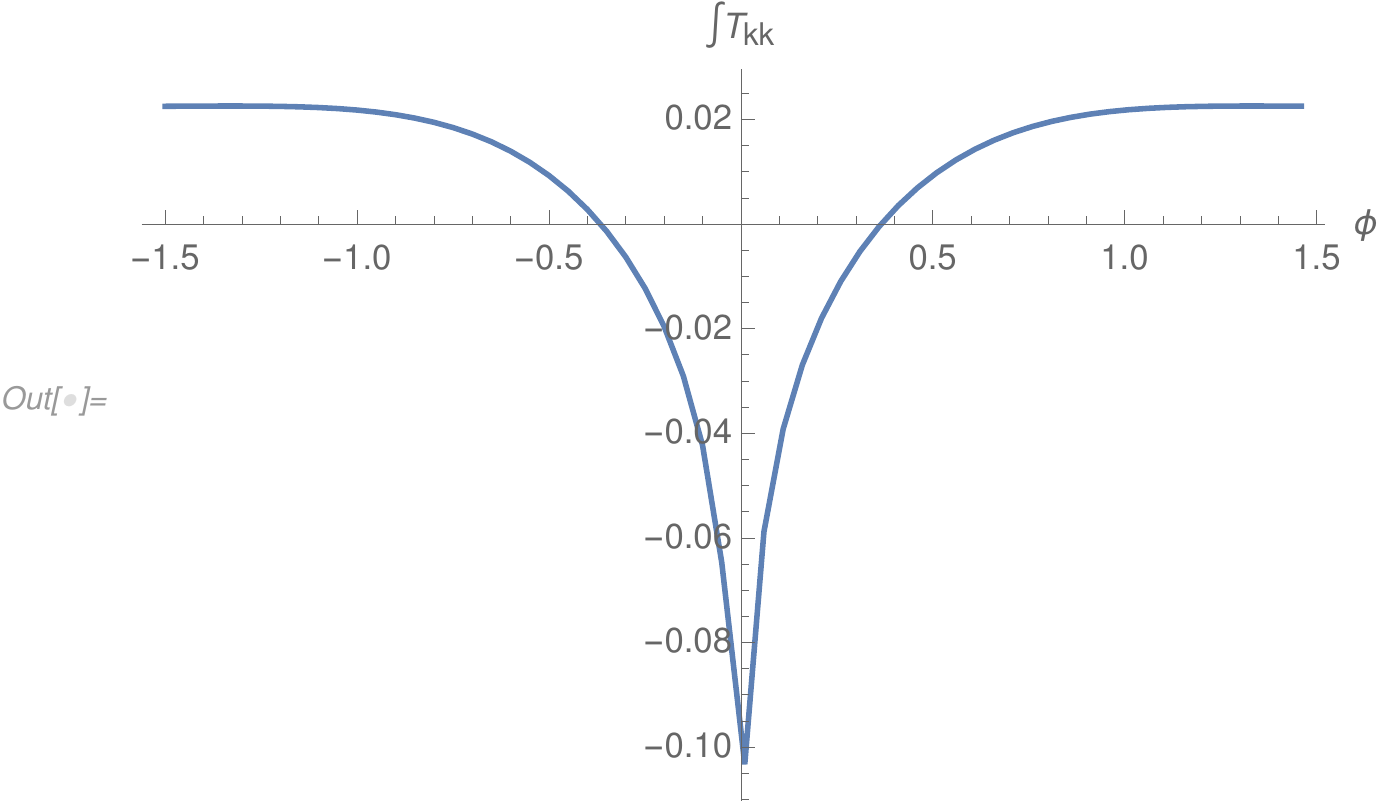}
\caption{A $\phi$ profile of $\int \text{d}U \langle T_{kk} \rangle$, which is negative for small $|\phi|$ and positive for larger $|\phi|$. $\Delta(p=0) = 1$, the $(-)$ sign is chosen for $p=1$, and the $(+)$ sign is chosen for all higher $p$. $m^2 = -0.2$, $\ell/R_{S^1} =0.1$, $r_+ = 1$ and $r_-=0$.}
\label{fig:changesign}
\end{figure}

In general, one finds $\int \text{d}U \langle T_{kk}\rangle$ to be negative for all $\phi$.  Positivity of the Green's function \eqref{eq:Hphi2} then shows that $\Delta V$ is negative at each $\phi$ and the wormhole is traversable when entered from any direction.  However, much as in section \ref{sec:EOTW}, one can engineer exceptions to this general rule by making use of the dependence of the integrals on $\Delta$.  In this way one can find examples where the sign of $\Delta V$ does in fact depend on $\phi$ and the wormhole is traversable only when entered from certain directions, see figure \ref{fig:changesign}.  The interesting feature of such examples is that they are then traversable with either periodic or anti-periodic boundary conditions, though the directions from which one must enter the wormhole to traverse it are complimentary in the two cases.

 \section{Discussion}
\label{sec:disc}

The above work studied back-reaction from quantum scalar fields in Hartle-Hawking states on simple explicit examples of ${\mathbb Z}_2$ wormholes asymptotic to AdS$_3$ and AdS$_3 \times S^1$.  These examples generally become traversable when the scalar satisfies periodic boundary conditions around the ${\mathbb Z}_2$ cycle, though as described in section \ref{sec:EOTW} one may engineer examples where this fails and anti-periodic boundary conditions are required for traversability. The examples of section \ref{sec:KKZBO} break rotational symmetry and, while they generally become traversable everywhere with periodic boundary conditions around the ${\mathbb Z}_2$-cycle, with care they can be similarly engineered to become traversable only for observers entering the wormhole at certain values of the angular coordinate $\phi$.

The most interesting result came from the rotating examples of section \ref{sec:KKZBO}, where we found the back-reaction to diverge when the background spacetime became extremal.  Though our analysis is perturbative, even when sourced by only a single scalar quantum field this suggests that a fully non-perturbative treatment would find a self-supporting eternal wormhole.  Indeed, the growth of our effect at small temperatures $T$
is directly analogous to the $\Delta > 1/2$ cases studied in \cite{Maldacena:2018lmt} where the perturbation grows in the IR limit.  Though the potential for diverging back-reaction at extremality was also simultaneously and independently found in \cite{Caceres:2018ehr}, such a divergence did not in fact arise in their context.

The diverging back-reaction near extremality follows directly from the linearized Einstein equations.  In our examples the extremal spacetimes are smooth and contain a non-contractible ${\mathbb Z}_2$ cycle of finite length.  As a result, it is natural in our examples (but in contrast to the setting studied in \cite{Caceres:2018ehr}) that $\int \text{d}U \langle T_{k k}\rangle$ remains non-zero and negative at extremality.  But from \eqref{eq:introt} any finite such perturbation causes a divergence in the zero-mode of the metric perturbation $h_{kk}$.  Thus the wormhole becomes traversable along each generator of the background horizon and -- at least at first order in perturbation theory -- the wormhole appears to remain open for arbitrarily long times as the extreme limit ($T=0$) is approached.  It would be useful to better understand the apparent lack of dependence on  $r_+=r_-$ in the resulting first-order $T\Delta V$ shown in figure \ref{fig:DVT}.

If this conjecture is correct, the breaking of rotational symmetry appears to play a key role in the construction.  In particular, we conjecture the existence of time-independent such wormholes with arbitrary size for the wormhole throat, and thus presumeably with arbitrary total mass.  Now, the attentive reader will notice that we have worked in what are effectively co-rotating coordinates.  So by `time-independent,' we mean invariant under translations along a co-rotating Killing.  And the lack of rotational symmetry means that our conjectured spacetimes should not be invariant under standard translations of the boundary time $t$.  This is important for consistency with the conjecture about arbitrary mass as (in the absence of horizons) Hamilton's equations imply that the generator of time-translation symmetry should be constant along any one-parameter family of time-independent solutions.  We thus expect that $M$ varies but $M-J$ is constant along our family of wormholes, and that (as one would also expect from supersymmetry considerations) even with quantum corrections the condition for extremality remains $M-J=0$.

While we have not performed a complete analysis of more general cases, and while the Aretakis instability is strongest for large angular momentum \cite{Casals:2016mel} and our instability appears to occur only for the zero mode while, it is nevertheless natural to expect our effect to be related to other known instabilities of extreme black holes \cite{Marolf:2010nd,Marolf:2011dj,Aretakis:2011ha,Aretakis:2011hc,Aretakis:2011gz,Aretakis:2012ei,Lucietti:2012sf} and thus to be generic in the extremal limit.  This may make the construction of self-supporting wormholes more straightforward than might otherwise be expected.

Indeed, as described in section \ref{sec:general} our basic framework applies much to much more general cases than those studied explicitly here.  Given any globally hyperbolic ${\mathbb Z}_2$ quotient of a spacetime with bifurcate Killing horizons and a well-defined Hartle-Hawking states under an isometry that exchanges the left-moving and right-moving horizons, at least one choice of boundary conditions (periodic or anti-periodic) for free scalar fields on that spacetime must give a (transient) traversable wormhole.  As described in appendix \ref{sec:stationary}, there may also be generalizations in which the covering space has no Killing symmetry and the horizon is merely stationary (i.e., both divergence-free and shear-free).

We have studied only scalar fields in detail, but the general arguments of section \ref{sec:disc} apply equally well to  higher spin fields. It would be especially interesting to study back-reaction from linearized gravitons, which are always present for spacetime dimension $d \ge 4$.  Indeed, they are in principle relevant even to our AdS$_3$ constructions that involve Kaluza-Klein directions (so that the full spacetime has $d\ge 4$).  Indeed, since in those examples the amount of negative energy is governed by the Kaluza-Klein scale, one expects contributions from gravitons to be similar to those of scalars despite the absence of 3-dimensional gravitons.  And since changing the sign of the metric perturbation is not a symmetry of the full Einstein-Hilbert theory, only periodic boundary conditions will be physically relevant.  One would generally expect gravitons to contribute with the same sign as other bosons, and in particular with the scalars studied above.  We therefore expect inclusion of gravitons to make our wormholes even more traversable.  Should this expectation turn out to be false, one could nevertheless ensure that the wormhole becomes traversable by adding an order one number of additional scalar fields.

\begin{figure}[t]
\begin{center}
\begin{subfigure}{.3\linewidth}
\includegraphics[width=0.9\textwidth]{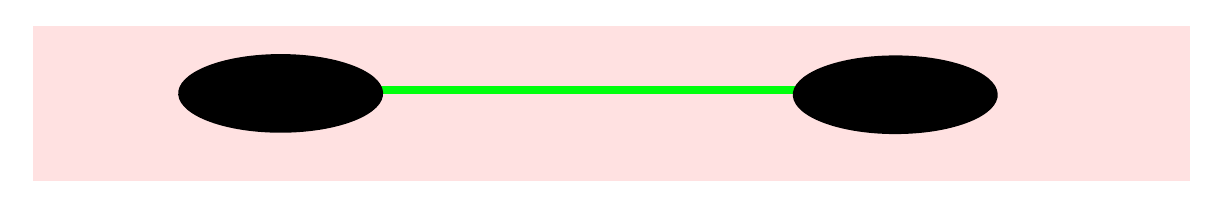}
\end{subfigure}
\begin{subfigure}{.3\linewidth}
\includegraphics[width=0.9\textwidth]{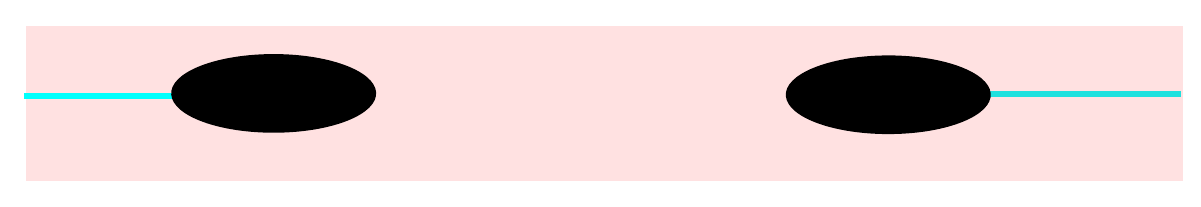}
\end{subfigure}
\begin{subfigure}{.3\linewidth}
\includegraphics[width=0.9\textwidth]{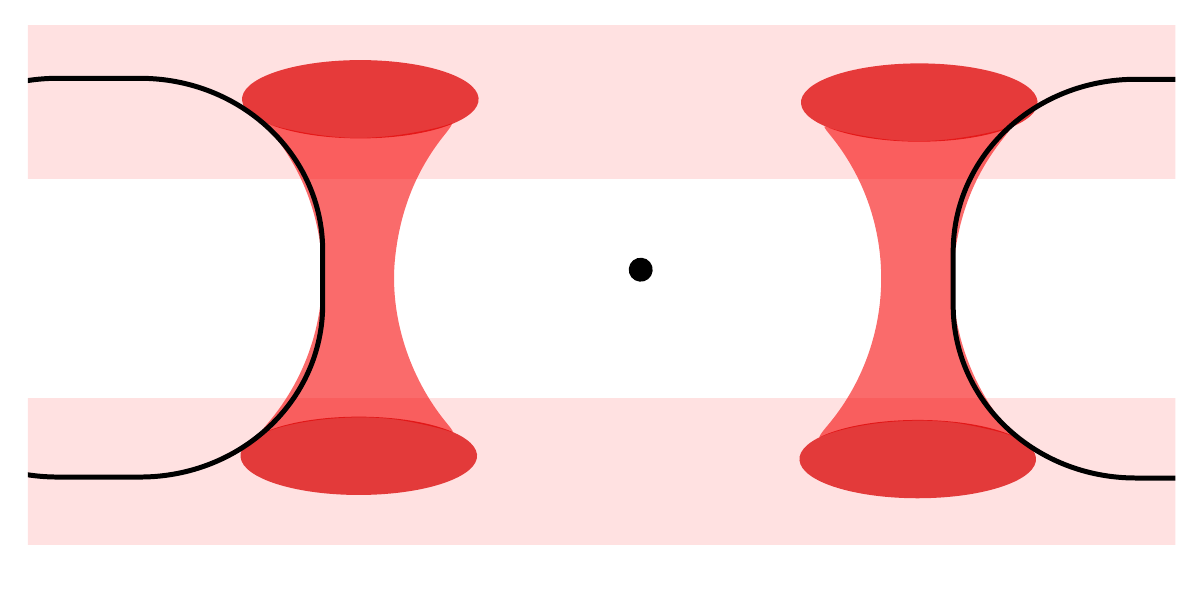}
\end{subfigure}
\end{center}
\caption{A moment of time in a spacetimes containing two black holes (black disks) held apart by a negative-tension strut (left) or by cosmic strings stretching to infinity (center and right) along the $z$-axis. The right-most figure shows both asymptotic regions and the wormholes that connect them.  The ${\mathbb Z}_2$ quotient described in the text acts as a $\pi$ rotation about the non-physical point indicated by the dot at the center of the right figure.}
        \label{fig:BachWeyl}
\end{figure}

While it is natural to think of the above quotients as geon-like (i.e., as generalizations of the $\mathbb{RP}^3$ geon described in \cite{GiuliniPhD,Friedman:1993ty}), they can also describe more familiar wormholes of the form shown in figure \ref{fig:wormhole} with wormhole homotopy group $\mathbb{Z}$. To see this, recall that static axisymmetric vacuum solutions to $d=4$ Einstein-Hilbert gravity take a simple form \cite{Weyl1917} found by Weyl in 1917, and that particular examples \cite{Bach1922} found by Bach and Weyl in 1922 can be understood \cite{Israel1964} as describing a pair of Schwarzschild black holes separated along the $z$-axis. The black holes are prevented from coalescing by a strut (i.e., by a negative tension cosmic string) along the axis between them and/or by positive-tension cosmic strings stretching from each black hole to infinity along the $z$-axis as shown in figure \ref{fig:BachWeyl}.  Furthermore, as described in \cite{Israel1964}, a natural analytic extension of this solution beyond the horizons gives a geometry with two asymptotically flat regions and a bifurcate Killing horizon.  The spacetime is thus similar to the standard Kruskal extension of the Schwarzschild black hole, except that this connection involves  a pair of wormholes (threaded by cosmic strings); see figure \ref{fig:BachWeyl} (right). This defines the ${\mathbb Z}_2$ cover $\tilde M$ of the desired spacetime $M$.

To construct $M$ itself, we simply note that $\tilde M$ has a ${\mathbb Z}_2$ symmetry $J$ that acts by simultaneously reflecting across the bifurcation surface and the surface $z=0$; i.e., it simultaneously exchanges the two sheets shown in figure \ref{fig:BachWeyl} (right) and also exchanges the two wormholes; i.e., it acts as a $\pi$ rotation about the non-physical point marked at the center of figure \ref{fig:BachWeyl} (right).  This $J$ has no fixed points, so $\tilde M = M/J$ is smooth up to cosmic strings and takes the familiar form described by figure \ref{fig:wormhole}.

In fact, at least in the positive-tension case, much as in section \ref{sec:KKZBO} it is straightforward to go one step farther and describe $\tilde M$ as the Kaluza-Klein reduction of a completely smooth spacetime.  Here one simply chooses parameters so that the cosmic strings are associated with deficit angles $2\pi(1-1/n)$.  We then consider a 5-dimensional spacetime $\tilde M_{KK}$ that is just $\tilde M \times S^1$ away from the strings.  At the location of the 4-dimensional cosmic strings, we instead take $M_{KK}$ to be locally what one might call the Kaluza-Klein cosmic string defined by $M^{3,1} \times S^1/\mathbb{Z}_n$ with the ${\mathbb Z}_n$ isometry acting by simultaneous rotations by $2\pi/n$ along the $S^1$ and about the $z$-axis\footnote{This 5-dimensional spacetime is usually Kaluza-Klein reduced along a different Killing field and then interpreted as a 4-dimensional spacetime sourced by a magnetic field \cite{Dowker:1993bt,Dowker:1994up}.  Since the energy of the solution is fixed by Noether's theorem independent of the reduction, the results of \cite{Dowker:1993bt,Dowker:1994up} show that the reduction used here gives a 4d solution with positive tension (rescaled from \cite{Dowker:1993bt,Dowker:1994up} by the relative length of their Kaluza-Klein circle relative to ours) but which in our case is 4d vacuum except at the string singularity on the $z$-axis.}.  This $M_{KK}$ is then a smooth ${\mathbb Z}_2$ quotient of a 5d spacetime ${\tilde M}_{KK}$ with bifurcate Killing horizon.  Since the spacetime is static and smooth, it also supports a Hartle-Hawking state defined by the Euclidean path integral.  Thus the analysis of section \ref{sec:general} applies and -- barring a miraculous general cancellation -- at least for generic values of parameters the wormhole must become traversable under first-order back-reaction from either periodic or anti-periodic scalar fields\footnote{Indeed, since this example breaks rotational symmetry it may be that both cases become traversable, with traversability being achieved along different generators for each of the two boundary conditions.}.

Although the form of the metric becomes more complicated, one may also add electric charge to the above solution as described in \cite{Stephani:2003tm}.  This would then provide an example of the standard wormhole form shown in figure \ref{fig:wormhole} with a smooth extremal limit satisfying all requirements from section \ref{sec:general} and in particular admitting a well-defined Hartle-Hawking state. In contrast, even at extremality, the rotating version will spin down due to spontaneous emission of angular momentum via the super-radiant modes \cite{Page:1976ki}, though this effect will in practice be slow for large black hole.

It would be interesting to analyze such examples in more detail, especially in the extreme limit.  Here the non-contractible cycles become long in the extreme limit, so that $\int \text{d}U \langle T_{kk}\rangle$ may become vanishingly small.  But the instability of extreme black holes raises the hope that even a vanishingly small perturbation could render the wormhole self-supporting and eternal at zero temperature.  Indeed, a naive analysis ignoring the redshift and issues associated with normalizing the affine parameter along the horizon would note that the length of a Reissner-Nordstr\"om throat grows like $T^{-1/2}$ so that an integrated Casmir-like energy would decay as $T^{1/2}$.  An instability that grows like $T^{-1}$ as in \eqref{eq:introt} would then suggest an eternal self-supporting wormhole. We will perform a more complete analysis using an effective 2-dimensional description for a model with conformal invariance in the near future.  If a large back-reaction does result, it would provide a simple perspective explaining the existence of the self-supporting wormhole recently constructed in \cite{Maldacena:2018gjk} -- here with the wormhole mouths kept from coalescing by cosmic strings instead of the orbital angular momentum used in \cite{Maldacena:2018gjk}.


\section*{Acknowledgements}
It is a pleasure to thank Ahmed Almheiri, Jorma Louko, Jim Hartle, Gary Horowitz, Juan Maldacena, Alexandros Mousatov, Jorge Santos, Milind Shyani, Xiaoliang Qi, and Aron Wall for useful discussions.  DM was supported in part by the U.S. National Science Foundation under grant number PHY15-04541 and by the University of California.   The final portion of his work was performed at the Aspen Center for Physics, which is supported by National Science Foundation grant PHY-1607611.  ZF was supported in part by the University of California.  BG-W was supported by a National Science Graduate Foundation Research Fellowship.

\appendix

\section{First-order traversability requires a stationary horizon}
\label{sec:stationary}

We show here that any background spacetime obeying the null convergence condition $R_{ab}k^ak^b \ge 0$ which can yield a traversable wormhole after first-order backreaction of a quantum field must be a quotient of a spacetime with a stationary (divergence-free and shear-free) horizon.

We phrase the argument for a spacetime $\tilde m$ with a single boundary\footnote{We use this term to refer to the regular part of the boundary; i.e., the part that is asymptotically flat or AdS and not the part of the conformal boundary describing spacetime singularities.}, but the argument for multiple boundaries is identical. Consider any curve that starts and ends at the boundary but is not smoothly deformable (with fixed endpoints) to lie entirely in the boundary. Let us now deform this curve by moving one endpoint to the far future on the boundary and the other to the far past on the boundary.  If the limiting curve could be causal with any timelike segment, there would be a faster causal curve through the wormhole (i.e., not deformable to lie in the boundary) which starts and ends on the boundary at finite times.  This is impossible since the wormhole is not traversable in the background \cite{Friedman:1993ty,Galloway:1999bp}).

Consider then the class of limiting curves that consist only of null and spacelike segments.  If the proper length of all such curves is bounded below, then no such curve can be rendered causal by an arbitrarily small perturbation.  Allowing timelike segments does not help, as that will necessarily make the spacelike segments longer.  So if the wormhole can be rendered traversable by an arbitrarily small perturbation, there must be a sequence of such limiting curves whose proper length approaches zero.  The limiting of this sequence is then a curve that is everywhere null.  (We assume the spacetime to be sufficiently regular so that this sequence is guaranteed to converge.)  It must also be a geodesic, else there would be a timelike curve that traverses the wormhole.  And since it runs from the boundary to the boundary, it is a complete null curve (having infinite affine parameter).

Now, since the spacetime contains a wormhole, it has some non-trivial wormhole homotopy group (see footnote \ref{foot:wpi}) that we can use to define a multiple cover $M$ of the original spacetime $\tilde M$.  The order of this cover does not matter.  In the cover, our complete null curve lifts to at least one complete null curve that starts one connected component of the boundary and ends on another.  That curve must be achronal, else the two boundaries would be causally connected (violating topological censorship  \cite{Friedman:1993ty,Galloway:1999bp}).  But since the original background (and thus the covering space) satisfies the null convergence condition, Galloway's splitting theorem (theorem 4.1 of \cite{Galloway:1999ny}) requires the geodesic to lie on a stationary null surface.  The projection of this surface to the original spacetime (a quotient of the cover) is thus stationary and null as well.

\bibliographystyle{utcaps}
\bibliography{wormholes}

\end{document}